\documentclass[11pt]{article}
\usepackage{fullpage}
\usepackage{amsfonts, amssymb, amsmath, amsthm}
\usepackage{latexsym}
\usepackage{url}
\usepackage{color}
\definecolor{DarkGray}{rgb}{0.1,0.1,0.5}
\usepackage[colorlinks=true,breaklinks, linkcolor= DarkGray,citecolor=DarkGray,urlcolor= DarkGray]{hyperref}	
\usepackage{graphicx}
\usepackage{subfigure}
\usepackage{cancel}
\usepackage{multirow}
\usepackage{caption}	

\usepackage{import}
\usepackage{ltxtable}

\usepackage{calc}	
\def\ket #1{\vert #1\rangle}
\def\bra #1{\langle #1\vert}
\def\ketbra #1#2{\ket{#1}\!\bra{#2}}

\def\abs #1{\lvert #1\rvert}
\DeclareMathOperator{\tr}{Tr}
\newcommand{\beq}{\begin{equation}}
\newcommand{\eeq}{\end{equation}}

\usepackage{color}
\newcommand{\comment}[1]{\emph{\color{red}Comment:\color{black} #1}} 
\newlength{\commentslength}
\newcommand{\comments}[1]{
\hspace{-2\parindent}
\addtolength{\commentslength}{-\commentslength}
\addtolength{\commentslength}{\linewidth}
\addtolength{\commentslength}{-\parindent}
\fcolorbox{red}{white}{\smallskip\begin{minipage}[c]{\commentslength}
\emph{Comments:}\begin{itemize}#1\end{itemize}\end{minipage}}\bigskip
}
\renewcommand{\comment}[1]{}\renewcommand{\comments}[1]{}

\newtheorem{theorem}{Theorem}  

\newtheorem{corollary}{Corollary}

\newtheorem{lemma}{Lemma}

\newtheoremstyle{note}{}{}{\slshape}{}{\bfseries}{.}{ }{}
\theoremstyle{note}

\newtheorem{definition}{Definition}

\newcommand{\eqnref}[1]{\hyperref[#1]{{(\ref*{#1})}}}
\newcommand{\thmref}[1]{\hyperref[#1]{{Theorem~\ref*{#1}}}}
\newcommand{\lemref}[1]{\hyperref[#1]{{Lemma~\ref*{#1}}}}
\newcommand{\corref}[1]{\hyperref[#1]{{Corollary~\ref*{#1}}}}
\newcommand{\defref}[1]{\hyperref[#1]{{Definition~\ref*{#1}}}}
\newcommand{\secref}[1]{\hyperref[#1]{{Section~\ref*{#1}}}}
\newcommand{\figref}[1]{\hyperref[#1]{{Figure~\ref*{#1}}}}
\newcommand{\tabref}[1]{\hyperref[#1]{{Table~\ref*{#1}}}}
\newcommand{\remref}[1]{\hyperref[#1]{{Remark~\ref*{#1}}}}
\newcommand{\appref}[1]{\hyperref[#1]{{Appendix~\ref*{#1}}}}
\newcommand{\claimref}[1]{\hyperref[#1]{{Claim~\ref*{#1}}}}
\newcommand{\exampleref}[1]{\hyperref[#1]{{Example~\ref*{#1}}}}

\DeclareMathOperator{\poly}{\operatorname{poly}}
\newcommand{\BQP}{{\mathsf{BQP}}}

\begin{document}

\author{%
Ben W.~Reichardt%
  \thanks{School of Computer Science and Institute for Quantum Computing, University of Waterloo.}}
\date{}

\def\magic #1{\mathcal{U}(#1)}

\title{Quantum universality by state distillation}

\date{}

\maketitle

\begin{abstract}
Quantum universality can be achieved using classically controlled stabilizer operations and repeated preparation of certain ancilla states.  Which ancilla states suffice for universality?  This ``magic states distillation" question is closely related to quantum fault tolerance.  Lower bounds on the noise tolerable on the ancilla help give lower bounds on the tolerable noise rate threshold for fault-tolerant computation.  Upper bounds show the limits of threshold upper-bound arguments based on the Gottesman-Knill theorem.  

We extend the range of single-qubit mixed states that are known to give universality, by using a simple parity-checking operation.  For applications to proving threshold lower bounds, certain practical stability characteristics are often required, and we also show a stable distillation procedure.  

No distillation upper bounds are known beyond those given by the Gottesman-Knill theorem.  One might ask whether distillation upper bounds reduce to upper bounds for single-qubit ancilla states.  For multi-qubit pure states and previously considered two-qubit ancilla states, the answer is yes.  However, we exhibit two-qubit mixed states that are not mixtures of stabilizer states, but for which every postselected stabilizer reduction from two qubits to one outputs a mixture of stabilizer states.  Distilling such states would require true multi-qubit state distillation methods.
\end{abstract}

\section{Introduction} \label{s:introduction}

Stabilizer operations, consisting of Clifford group unitaries and Pauli operator measurement and eigenstate preparation, suffice for generating interesting, highly entangled quantum states.  By the Gottesman-Knill theorem, however, they are efficiently classically simulatable and not quantum universal~\cite{AaronsonGottesman04}.  What more does one need to obtain quantum universality?  A sufficient additional operation is any one-qubit unitary that is not a Clifford up to overall phase~\cite{Shi02}.  Almost every $n$-qubit unitary also suffices.  Much less is known, however, about which non-unitary quantum channels, such as noisy gates, suffice for universality together with stabilizer operations.  If we assume that the quantum operations are under the adaptive control of a universal classical computer, then this question turns out to be a special case of a broader problem: 
 
\newtheorem*{mstatesdistillationproblem}{Magic states distillation problem}
\begin{mstatesdistillationproblem}
For which quantum states $\rho$ does stabilizer operations plus repeated preparation of $\rho$ imply quantum universality?
\end{mstatesdistillationproblem}

If repeated preparation of $\rho$ and stabilizer operations gives universality, we say for short $\magic{\rho}$.  The problem of characterizing the states $\rho$ for which $\magic \rho$ holds has its main application in quantum fault tolerance.  For fault-tolerance schemes based on stabilizer codes, encoded stabilizer operations are the easiest operations to implement.  Magic states distillation allows extending these operations to a full universal set, provided that noisy enough states $\rho$ satisfy $\magic \rho$.  Conversely, limits on when $\magic \rho$ can hold can also give limits on fault-tolerance schemes.  

In this paper, we extend the range of single-qubit mixed states $\rho$ that are known to give universality, by using a simple parity-checking operation.  However, the question of fully characterizing those single-qubit states $\rho$ for which $\magic \rho$ holds remains open.  For multi-qubit pure states, the question $\magic \rho$?\ reduces to the same question for single-qubit pure states.  Unfortunately, though, as a second result we show that this question for mixed multi-qubit states $\rho$ generally does \emph{not} reduce to the single-qubit problem.  We also study the applications of magic states distillation to quantum fault-tolerance schemes.  In particular, these applications in practice will require certain stability properties, and we present a stable distillation procedure.

\subsection{Single-qubit magic states distillation}

\begin{figure}
\begin{center}
\includegraphics[scale=.3928]{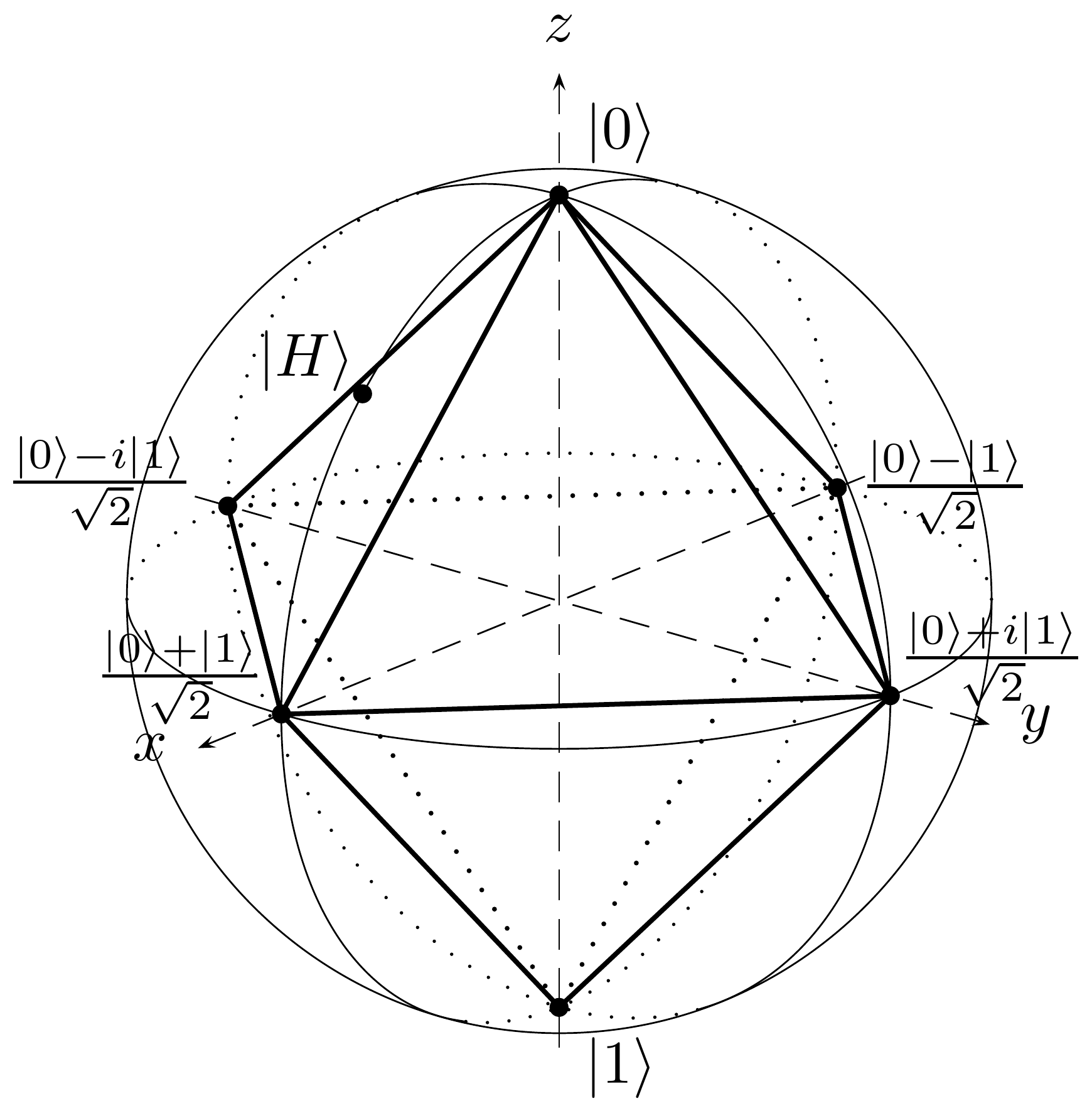}
\end{center}
\caption{{Bloch sphere:} 
Up to a phase, single-qubit states are in correspondence with points on or in the unit sphere in ${\bf R}^3$.  The state $\rho(x,y,z) = \tfrac{1}{2}(I + x X + y Y + z Z)$ corresponds to the point $(x,y,z)$.  Pure states correspond to points on the surface of the sphere.  All points $\rho$ in the octahedron $\mathcal{O}$ the convex hull of the six Pauli eigenstates can be prepared using stabilizer operations.} \label{f:blochsphere}
\end{figure}

Bravyi and Kitaev~\cite{BravyiKitaev04} posed and studied the magic states distillation problem for single-qubit states $\rho$.  Single-qubit states are conveniently parametrized by the Bloch sphere of \figref{f:blochsphere}, under the correspondence $(x, y, z) \leftrightarrow \rho(x,y,z) = \tfrac{1}{2} (I + x X + y Y + z Z)$.  Here $X = \left(\begin{smallmatrix}0&1\\1&0\end{smallmatrix}\right)$, $Y = \left(\begin{smallmatrix}0&-i\\i&0\end{smallmatrix}\right)$, $Z = \left(\begin{smallmatrix}1&0\\0&-1\end{smallmatrix}\right)$ are the Pauli matrices.  Bravyi and Kitaev proved that certain $\rho$ can be efficiently distilled with stabilizer operations to be arbitrarily close to the state $\ket H := \cos\frac{\pi}8 \ket 0 + \sin\frac{\pi}8 \ket 1 \leftrightarrow (\frac1{\sqrt 2}, \frac1{\sqrt 2}, 0)$.  Knill, Laflamme and Zurek~\cite{KnillLaflammeZurekProcRSocLondA98} proved $\magic{\ket H}$.  Combining these results gives: 

\begin{theorem}[\cite{BravyiKitaev04}] \label{t:BravyiKitaev04}
$\magic{\rho}$ holds for single-qubit states $\rho(x,y,z)$ with 
\begin{align}
&& \max\{\abs{x} + \abs{z}, \abs{x} + \abs{y}, \abs{y} + \abs{z}\}  &> 1.015 \label{e:BravyiKitaev04H} \\ 
&\text{or} & \abs{x} + \abs{y} + \abs{z} &> 3/\sqrt{7} \approx 1.134
 \enspace . \label{e:BravyiKitaev04T}
\end{align}
\end{theorem}

On the other hand, it is obvious that $\magic{\rho}$ does not hold for $\rho(x,y,z)$ with $\abs x + \abs y + \abs z \leq 1$, as such states are convex combinations of the Pauli eigenstates $\{(\pm 1, 0, 0), (0, \pm 1, 0), (0,  0, \pm 1)\}$.  Such states we call ``stabilizer states," since they can be prepared using stabilizer operations.  

Thus \thmref{t:BravyiKitaev04} leaves a gap in between the stabilizer states and the known distillable states.  What happens in the region between them?  Non-adaptive stabilizer operations alone compute the class $\mathsf{\oplus L}$, so are probably not universal even for classical computation~\cite{AaronsonGottesman04}.  One intriguing possibility is that stabilizer operations with states $\rho(x,y,z)$ outside the region of \thmref{t:BravyiKitaev04} but with $\abs x + \abs y + \abs z > 1$ could give an intermediate class between $\mathsf{BPP}$ and $\BQP$.  Another possibility is that there is a sharp threshold, i.e., that $\magic{\rho(x,y,z)}$ holds exactly when $\abs x + \abs y + \abs z > 1$.  

In fact, Ref.~\cite{Reichardt04magic} showed that there is indeed a sharp threshold in the $xz$-plane of the Bloch sphere:

\begin{theorem}[\cite{Reichardt04magic}] \label{t:Reichardt04magic}
$\magic{\rho(x,y,z)}$ holds if 
\begin{equation} \label{e:Reichardt04magicH}
\max\{\abs{x} + \abs{z}, \abs{x} + \abs{y}, \abs{y} + \abs{z}\}  > 1
 \enspace .
\end{equation}
\end{theorem}

The improvement of Eq.~\eqnref{e:Reichardt04magicH} over Eq.~\eqnref{e:BravyiKitaev04H} is tight, as the states $\rho(x, 0, 1-x)$ are stabilizer states.  \thmref{t:Reichardt04magic} also implies:

\begin{corollary}[{\cite{Reichardt04magic}}] \label{t:Reichardt04magicpure}
$\magic{\rho}$ holds for every single-qubit pure state $\rho$ that is not one of the six Pauli eigenstates.  
\end{corollary}

\noindent Pure states correspond to points on the surface of the Bloch sphere, i.e.,  $x^2 + y^2 + z^2 = 1$.

In this paper, we extend the set of single-qubit states $\rho$ for which we know $\magic{\rho}$ slightly further: 

\begin{theorem}
 \label{t:rotations}
$\magic{\rho(x,y,z)}$ holds if 
\begin{equation} \label{e:rotations}
\max\{ \abs{x} + \sqrt{y^2+z^2}, \abs{y} + \sqrt{x^2+z^2}, \abs{z} + \sqrt{x^2+y^2}\} > 1
 \enspace .
\end{equation}
\end{theorem}

The basic operation required is a simple parity check, which we introduce in \secref{s:universality}.  Applying the parity check in the computational and dual bases, in \secref{s:fourqubitcode}, gives the stated improvement in the distillable region of states.  

Moreover, we show in \secref{s:fivequbit} that the set of distillable states is strictly larger than the set delimited by Eqs.~\eqnref{e:BravyiKitaev04T} and~\eqnref{e:rotations}: 

\begin{theorem} \label{t:fivequbittwisted}
$\magic{\rho(f x, f y, f z)}$ holds for $x = y = \tfrac{3 \sqrt{7} - 7}{7 (2 - \sqrt{2})} \approx 0.229$, $z = \tfrac{14 - 3 \sqrt{14}}{7 (2 - \sqrt{2})} \approx 0.677$ and $f = 0.9895$.
\end{theorem}

Notice that the values of $(x, y, z)$ in \thmref{t:fivequbittwisted} satisfy Eqs.~\eqnref{e:BravyiKitaev04T} and~\eqnref{e:rotations} with equality, so $(fx,fy,fz)$ is a slight but strict improvement.  The proof of \thmref{t:fivequbittwisted} comes from a small modification of Bravyi and Kitaev's distillation scheme.  

\begin{figure}
\centering
\begin{tabular}{c@{$\quad$}c}
$\qquad \qquad \qquad$\subfigure[]{\label{f:oldnorotation}\includegraphics*[scale=.5]{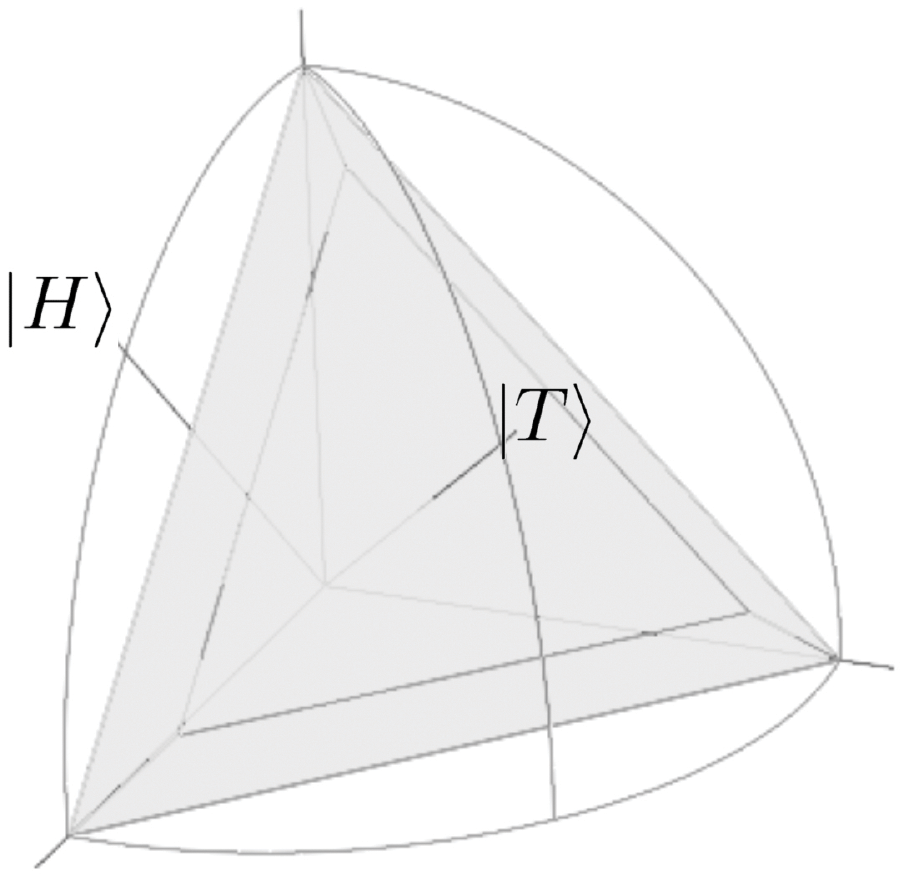}}&
\subfigure[]{\label{f:rotationsandtsmall}\includegraphics*[scale=.5]{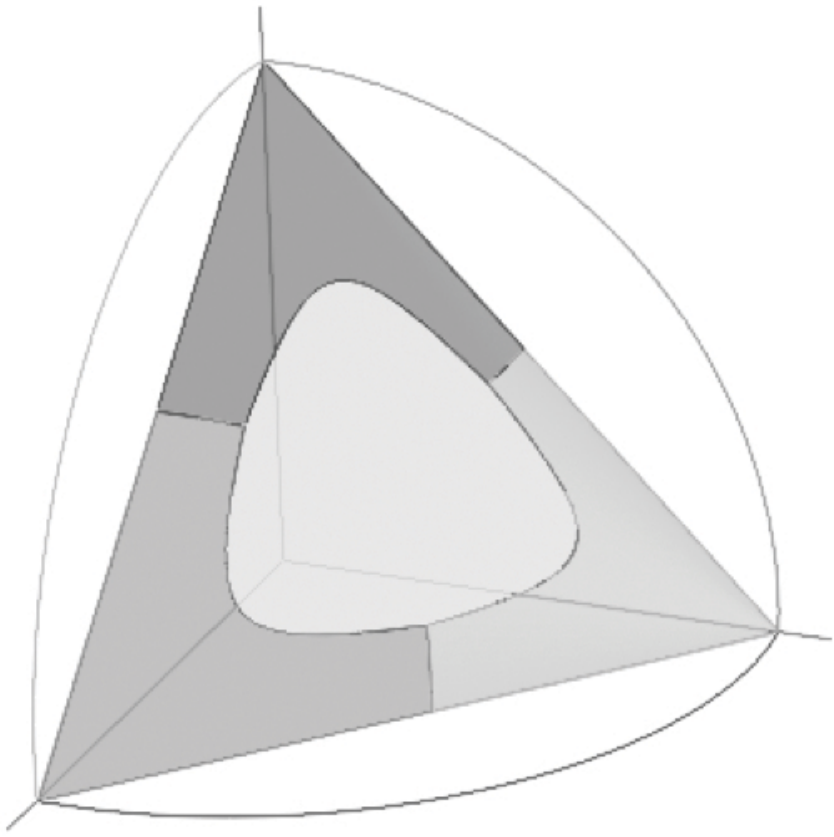}}\\
\\\multicolumn{2}{c}{\subfigure[]{\label{f:Graph_labeled}\includegraphics[scale=.58]{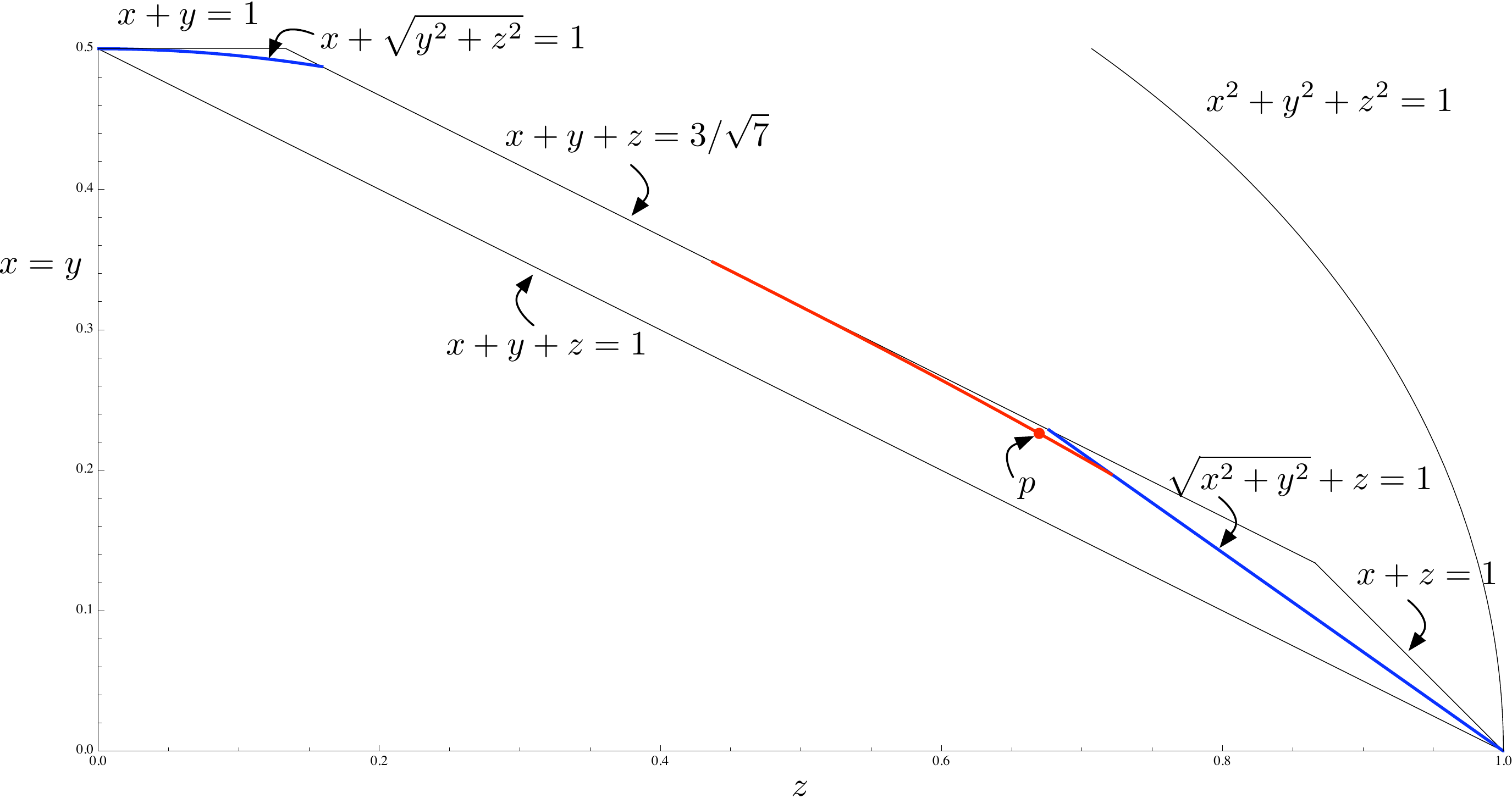}}}
\end{tabular}
\caption{{Open region:} In (a) is shown the region of single-qubit states for which $\magic{\rho}$ had not been known, bounded in one octant of the Bloch sphere by $1 < x+y+z \leq 3/\sqrt{7}$ and $\max\{x+y, y+z, x+z\} \leq 1$.  The other octants are symmetrical.  The region remaining after \thmref{t:rotations}, which adds the inequality $\max\{ x + \sqrt{y^2+z^2}, y + \sqrt{x^2+z^2}, z + \sqrt{x^2+y^2}\} \leq 1$, is shown in (b).  Part~(c) shows a cross-section of this octant through the plane $x = y$.  The narrowings of the open region due to \thmref{t:rotations} are indicated with thick blue curves.  The thick red curve shows the improvement from the distillation procedure of \thmref{t:fivequbittwisted}; the point $p = 0.9895 ( \tfrac{3 \sqrt{7} - 7}{7 (2 - \sqrt{2})}, \tfrac{3 \sqrt{7} - 7}{7 (2 - \sqrt{2})}, \tfrac{14 - 3 \sqrt{14}}{7 (2 - \sqrt{2})})$ is indicated.  This curve has been computed numerically, and we do not know a parametric form.} \label{f:oldnorotation_rotationsandt}
\end{figure}

\figref{f:oldnorotation_rotationsandt} displays the new distillable state results.  

\subsection{Multi-qubit state distillation}

In~\cite{Reichardt04magic}, this author considered the question $\magic{\rho}?$ for multi-qubit pure states.  For every multi-qubit pure state that is not a stabilizer state, there exists a sequence of stabilizer operations that reduces the state down to a single-qubit pure state that is not a Pauli eigenstate.  From \corref{t:Reichardt04magicpure}, this implies: 

\begin{corollary}[{\cite{Reichardt04magic}}] \label{t:Reichardt04magicpuremulti}
$\magic{\rho}$ holds for every single- or multi-qubit pure state that is not a stabilizer state.
\end{corollary}

\begin{corollary} \label{t:Reichardt04magicunitarymulti}
Almost every $n$-qubit unitary, together with stabilizer operations, gives quantum universality.  
\end{corollary}

\noindent \corref{t:Reichardt04magicunitarymulti} follows from \corref{t:Reichardt04magicpuremulti} by applying the unitary to $\ket{0^n}$; the result will almost certainly not be a stabilizer state, since there are a finite number of pure $n$-qubit stabilizer states.  

We here study distillation of two-qubit mixed states.  We provide an example of a two-qubit state that is not a mixture of stabilizer states, but for which every two-to-one-qubit stabilizer reduction outputs a mixture of stabilizer states (\secref{s:twoqubitstate}).  This implies that, unlike the situation for pure states, the question of $\magic{\rho}?$ for mixed states does not reduce to the single-qubit case.  

\begin{theorem} \label{t:twoqubitcounterexample}
The state $\rho = \tfrac{1}{4}II + \tfrac{1}{12}(IY + IZ - XX + YX + ZX)$ is not a mixture of stabilizer states, but every two-to-one-qubit stabilizer reduction outputs a mixture of single-qubit stabilizer states.
\end{theorem}

\thmref{t:twoqubitcounterexample} is proved by studying the $15$-dimensional polytope of two-qubit stabilizer states.  However, the problem of multi-qubit magic states distillation is still mostly open; for example, we do not know if $\magic{\rho}$ holds for the $\rho$ of \thmref{t:twoqubitcounterexample}, nor do we know of any two-qubit distillation procedure that cannot start with reduction to a one-qubit state.  Dennis has previously considered distillation of a two-qubit state for universality~\cite{Dennis01distill}.  However, for the state, $\tfrac{1}{\sqrt{3}}(\ket{00}+\ket{01}+\ket{10})$, and noise model in \cite{Dennis01distill}, as well as for simultaneous depolarizing noise, universality does reduce to the single-qubit case.  Below the noise level at which the state becomes a mixture of stabilizer states, universality can be obtained by measuring the second qubit to be $\ket{+}$ and applying \thmref{t:Reichardt04magic}.

\subsection{State distillation and quantum fault tolerance}

This magic states distillation problem is natural from a quantum information point of view, with the motivation coming from understanding the gap between classically controlled stabilizer operations ($\mathsf{BPP}$) and full quantum universality ($\BQP$).  The main application, though, is to lower and upper bounds for quantum fault tolerance (\secref{s:faulttolerance}).  

Magic states distillation serves as a reduction from universal fault-tolerant computation down to fault-tolerant stabilizer operations.  This reduction often works in practice without affecting the maximum tolerable noise rate, or \emph{threshold}, because the bottleneck is in achieving reliable stabilizer operations.  

Determining upper bounds on the tolerable noise threshold is a difficult problem.  We show that two recent upper bounds~\cite{VirmaniHuelgaPlenio,BuhrmanCleveLaurentLindenSchrijverUnger06}, based on upper-bounding the amount of noise before a gate set becomes classically simulatable by the Gottesman-Knill theorem, are unfortunately tight.  For example, 

\begin{theorem} \label{t:pipereighttheorem}
Classically controlled stabilizer operations, together with repeated application of a $\pi/8$ gate ($\exp(i \tfrac{\pi}{8}Z)$) subject to worst-case probabilistic noise at rate $p$ (or dephased at twice that rate), give universality if and only if $p < \tfrac{1}{\sqrt{2}}(1 - \tfrac{1}{\sqrt{2}})$.

Classically controlled stabilizer operations, together with repeated application of a $\pi/8$ gate depolarized at rate $p$, give universality if and only if $p < (6-2\sqrt{2})/7$.
\end{theorem}

The two ``only if" parts of \thmref{t:pipereighttheorem} are proved by~\cite{VirmaniHuelgaPlenio} and~\cite{BuhrmanCleveLaurentLindenSchrijverUnger06}, respectively.  The first ``if" part is a consequence of \thmref{t:Reichardt04magic}.  We prove the second ``if" part by distilling a certain \emph{two}-qubit state, based on the Jamiolkowski isomorphism.

\section{Universality of stabilizer operations and preparation of $\ket H$} \label{s:universality}

This section reviews the proof from Refs.~\cite{KnillLaflammeZurekProcRSocLondA98, BravyiKitaev04} of $\magic{\ket H}$ and introduces the parity-checking operation used in \secref{s:fourqubitcode} to implement an improved magic states distillation procedure.  This scheme is similar to the partner-pairing algorithm used in heat-bath algorithmic cooling~\cite{fernandez-2004-2}.

\begin{definition}
Stabilizer operations consist of Clifford group unitaries, preparation of $\ket{0}$ and measurement in the computational $\ket{0}, \ket{1}$ basis.  Clifford group unitaries are generated by the Hadamard gate $H = \tfrac{1}{\sqrt{2}}\left(\begin{smallmatrix}1&1\\1&-1\end{smallmatrix}\right)$, the phase gate $Z^{1/2} = \left(\begin{smallmatrix}1&0\\0&i\end{smallmatrix}\right)$ and the controlled-NOT gate, CNOT $\ket{a} \ket{b} = \ket{a} \ket{a + b \mod 2}$ for $a, b \in \{0,1\}$.  
\end{definition}

The CNOT gate and arbitrary single-qubit rotations from $SU(2)$ give universality~\cite{BarencoBennettCleveDiVincenzoMargolusShorSleatorSmolinWeinfurter95}, from which Boykin et al.\ proved the universality of CNOT, Hadamard, and $Z^{1/4}$~\cite{BoykinMorPulverRoychowdhuryVatan00}.  The intuition is that using $Z^{1/4}$ and its stabilizer conjugates, like $X^{1/4}$, it is easy to obtain a rotation by an irrational multiple of $\pi$ about some axis, and hence a dense set of rotations about that axis.  Rotations about a second axis can be obtained by conjugation with a stabilizer operation.

The proof of $\magic{\ket H}$ follows from a trick to implement $\exp(i \tfrac{\phi}{2} Z)$ on a data qubit using repeated preparation of $\exp(i \tfrac{\phi}{2} Z) \ket{+}$ and stabilizer operations with adaptive classical control.  Here $\ket{+} = \tfrac{1}{\sqrt{2}}(\ket{0}+\ket{1})$ is $+1$ eigenstate of $X$.

Single qubit pure states can be parameterized by polar coordinates $(\theta, \phi)$ on the Bloch sphere: $\ket{\psi(\theta,\phi)} = \cos \tfrac{\theta}{2} \ket{0} + e^{i \phi} \sin \tfrac{\theta}{2} \ket{1}$, so $\ket H = \ket{\psi(\pi/4, 0)}$.  Single qubit Clifford unitaries consist exactly of the rotational symmetries of the octahedron $\mathcal{O}$ of \figref{f:blochsphere}, so in particular $\ket{\psi(\theta,0)}$ and $\ket{\psi(\tfrac{\pi}{2},\theta)}$ are symmetrical.  Now notice that 
\begin{equation}
(\alpha \ket{0} + \beta \ket{1}) \otimes (\ket{0} + e^{i \theta} \ket{1}) = \alpha \ket{00} + \beta e^{i \theta} \ket{11} + \alpha e^{i \theta} \ket{01} + \beta \ket{10}
\end{equation}
by simply expanding out the tensor product.  Therefore, we can apply a CNOT between the qubits to measure the parity.  On measuring even parity, the operation $\left(\begin{smallmatrix}1&0\\0&e^{i \theta}\end{smallmatrix}\right)$ has been applied to the remaining qubit.  When the parity is odd, $\left(\begin{smallmatrix}1&0\\0&e^{-i \theta}\end{smallmatrix}\right)$ has been applied.  In the latter case, we repeat the process, carrying out a random walk on phases that are integer multiples of $\theta$.  Terminate the walk when the phase is $+\theta$.  Note the necessity of adaptive classical control of the quantum circuit.

This procedure in particular implements $Z^{1/4} \propto e^{i \tfrac{\pi}{8} Z}$ given repeated preparation of $\ket H$, giving quantum universality $\magic{\ket H}$.  In this case, a random walk is actually not necessary; on measuring odd parity, apply the correction $Y^{1/2}$.  In fact, Shi has extended~\cite{BarencoBennettCleveDiVincenzoMargolusShorSleatorSmolinWeinfurter95} to show universality of $\{CNOT, T\}$, where $T$ is any single-qubit real gate such that $T^2$ does not preserve the computational basis~\cite{Shi02}.  This implies $\magic{\ket{\psi(\theta,0)}}$ for almost all $\theta$.

\section{Parity-checking distillation algorithm} \label{s:fourqubitcode}

The same parity-checking operation can be applied to single-qubit mixed states.

\subsection{Case $y=0$}

First, assume that the $y$ coordinate is zero.  Taking two copies of $\rho(x,0,z)$, 
\beq
\rho(x,0,z) \otimes \rho(x,0,z)
= \tfrac{1}{4} \left(\begin{smallmatrix}
(1+z)^2 & x(1+z) & x(1+z) & x^2\\
x(1+z)   & 1-z^2 &  x^2    & x(1-z) \\
x(1+z)   & x^2   & 1-z^2 & x(1-z) \\
x^2      & x(1-z)  & x(1-z) & (1-z)^2
\end{smallmatrix}\right)
 \enspace .
\eeq
A successful parity check extracts the $2 \times 2$ submatrix of the first and last rows and columns.  Renormalizing this submatrix to have trace one, and converting back into $(x,y,z)$ Bloch sphere coordinates,
\beq \label{e:yzero}
(x,0,z) \mapsto (x',0,z') = (\tfrac{x^2}{1+z^2},0,\tfrac{2z}{1+z^2}) .
\eeq

Now assume additionally that $x=z$; this can be achieved by applying a Hadamard gate to $\rho$ with probability $1/2$.  Take two output qubits from two separate successful parity checks, and perform another parity check in the dual basis, i.e., switching $x$ and $z$.  Then after resymmetrizing about the $x=z$ axis, we get overall
\beq \label{e:yequalszerocase}
(x,0,x) \mapsto \tfrac{x^2(3+x^2)}{1+2x^2+2x^4} (1,0,1)
 \enspace .
\eeq
It is simple to verify that this function lies above $x$ for $1/2 < x \lesssim 0.68$.  Repeatedly apply this scheme to iterate Eq.~\eqnref{e:yequalszerocase}, in each step pairing together successful states from the previous iteration, and conditioning on success of all the measurements.  The states output by this procedure will not approach $\ket H$ (at $x = y = 1/\sqrt 2 \approx 0.71$).  However, since $2 \cdot 0.68 > 1.015$, we can use \thmref{t:BravyiKitaev04} to obtain universality.  This provides a new proof of \thmref{t:Reichardt04magic}.  

The procedure above is equivalent to taking four copies of $\rho$ and conditioning the state to lie in the logical subspace of the four-qubit erasure code.  In Ref.~\cite{Reichardt04magic}, \thmref{t:Reichardt04magic} is proved by using the seven-qubit Steane and 23-qubit Golay codes similarly.  This scheme is simpler, but less efficient and can only distill $\ket H$ indirectly.  

There are a various related distillation schemes.  For example, instead of dual-parity-checking two copies of the parity check's output, one can dual-parity-check $\rho$ and the parity check's output---equivalent to distillation with a certain \emph{three}-qubit code.  We have checked by exhaustive enumeration all three-to-one- and four-to-one-qubit distillation protocols, and these schemes appear to be optimal, insofar as they maximize $x'+z'$ starting from $x=z=\tfrac{1}{2}+0.001$.

\subsection{Case $y \neq 0$}

Next consider the case when initially $y \neq 0$.  Of course, symmetrizing about the Hadamard axis forces $y$ to 0, but we would like to do better.  With a single postselected parity check, 
\beq \label{e:ynonzerocase}
(x,y,z) \mapsto (x',y',z') = \tfrac{1}{1+z^2}(x^2-y^2, 2 x y, 2 z)
 \enspace .
\eeq
To understand Eq.~\eqnref{e:ynonzerocase}, reparametrize to $\big(r=\sqrt{x^2+y^2}, \phi = \arctan(y/x), z\big)$.  Then since 
\beq
y'/x' = \frac{2xy}{x^2-y^2} = \tan 2\phi
 \enspace ,
\eeq
a postselected parity check accomplishes the mapping 
\beq
(r, \phi, z) \mapsto (r', \phi', z') = (\tfrac{r^2}{1+z^2}, 2\phi, \tfrac{2z}{1+z^2})
 \enspace .
\eeq
Thus, $r$ and $z$ transform just as $x$ and $z$ do in the case $y=0$; but also the angle $\phi$ doubles.  Now if $\phi = k \pi / 2^l$ for integers $k$, $l$, then repeated doubling will eventually move the state into the $y = 0$ plane.  Now, as long as 
\beq
r + z  = \sqrt{x^2+y^2} + z > 1
 \enspace ,
\eeq
one can verify with straightforward calculus that 
\beq
r' + z' > 1
 \enspace .
\eeq
Therefore, once the state reaches the $y = 0$ plane, it will satisfy $x+z > 1$, so \thmref{t:Reichardt04magic} applies.  Unfortunately, this method is very inefficient, since the difference $r' + z' - 1$ can be quadratically smaller than $r + z -1$.  

For most states, of course, $\phi \notin \{ k \pi / 2^l : k, l \in \mathbf{Z}\}$.  However, since we are dealing with mixed states, there is a simple solution.  We can create any state $\rho_0$ within the octahedron $\mathcal{O}$, by preparing the Pauli eigenstates with the appropriate probabilities.  The set of ``nice" longitudinal angles $\{ k \pi / 2^l : k, l \in \mathbf{Z}\}$ is dense.  So we can certainly choose a $\rho_0$ (in fact, $\ket{+}$ will suffice) and mix it with $\rho$ with appropriate probabilities to generate $\rho'$ with a nice longitudal angle and such that the new state still has $r'+z'>1$.  This proves \thmref{t:rotations}.

\section{Five-qubit distillation procedures} \label{s:fivequbit}

In the proof of \thmref{t:BravyiKitaev04}, Bravyi and Kitaev use a $15$-qubit Reed-Muller code~\cite{KnillLaflammeZurek96} that allows transversal application of $\exp(i \tfrac{\pi}{8} Y)$ to prove universality when $\max\{\abs{x}+\abs{z},\abs{x}+\abs{y},\abs{y}+\abs{z}\} > 1.015$.  (This method is actually equivalent to a scheme given by Knill~\cite{Knill04schemes,Reichardt04magic}.)  

To prove $\magic{\rho(x,y,z)}$ when $\abs{x}+\abs{y}+\abs{z} > 3/\sqrt{7}$, they use stabilizer operations to project five copies of $\rho$ into the codespace of the five-qubit code, with stabilizer generators 
\beq
\begin{tabular}{
c @{\!\,} c @{\!\,} c @{\!\,} c @{\!\,} c
@{,\,}
c @{\!\,} c @{\!\,} c @{\!\,} c @{\!\,} c
}
X&Z&Z&X&I&
I&X&Z&Z&X, \\
X&I&X&Z&Z&
Z&X&I&X&Z.
\end{tabular}
\eeq
They then decode the logical qubit.  
\newcommand{\degree}{\ensuremath{^\circ}}
Repeatedly applying this procedure, the state converges to $\ketbra{T}{T} = \rho(\tfrac{1}{\sqrt{3}},\tfrac{1}{\sqrt{3}},\tfrac{1}{\sqrt{3}})$ as long as $x=y=z > 1/\sqrt{3}$.   Here $\ket T$ is the $e^{i 2\pi/3}$ eigenstate of $T = \tfrac{1}{2} \left(\begin{smallmatrix} -1+i & 1+i \\ -1+i & -1-i \end{smallmatrix}\right)$, a $120\degree$ rotation about $\frac{1}{\sqrt 3}(1,1,1)$ on the Bloch sphere.  In general of course $x$, $y$ and $z$ will be unequal.  In that case, first apply single-qubit unitaries to move $\rho$ into the positive octant, so $x,y,z>0$.  Then with equal probabilities $1/3$ apply either $I$, $T$ or $T^2$.  This symmetrizes $\rho$'s coordinates.

A simple modification of the above procedure can give improvements for asymmetric $\rho$.  For $x = y < z$, applying $T$ to two of the five copies of $\rho$ before distilling with the five-qubit code (without symmetrizing $\rho$'s coordinates) gives an improvement for $0 < x=y < 1/\sqrt{3}$, according to numerical iterations of the equation 
\begin{equation}
(x,y,z) \mapsto (x',y',z') = \frac{2}{\left(\begin{split}1+x^4+y^4+z^4\\+4 x y z(x+y+z)\end{split}\right)}\left(\begin{split}x^3-x^2(y^3+z^3)+yz(y+z+2x-xyz),\\-y^3+y^2(x^3+z^3)-xz(x+z+2y-xyz),\\z^3-z^2(x^3+y^3)+xy(x+y+2z-xyz)\end{split}\right)
\end{equation}
The thick red curve in \figref{f:Graph_labeled} shows the distillable region, computed numerically, using this procedure in the $x=y$ cross-section of the Bloch sphere.  There is an improvement between $x=y=0.1956$ and $x=y=z=1/\sqrt{7} \approx 0.378$.  In particular, \thmref{t:fivequbittwisted} is one special case.  

We have exhaustively searched all five-to-one-qubit stabilizer reductions, evaluating each on $\rho(\tfrac{1}{\sqrt{3}}(1,1,1))$ and $\rho(x=y=\tfrac{3 \sqrt{7} - 7}{7 (2 - \sqrt{2})}, z =  \tfrac{14 - 3 \sqrt{14}}{7 (2 - \sqrt{2})})$, and found no code performed better than the five-qubit code, either alone or with $T$ applied to one or two of the qubits.

\section{Distillation of multi-qubit states} \label{s:twoqubitstate}

The question $\magic{\ket{\psi}}?$ for multi-qubit pure states $\ket{\psi}$ reduces to the same question for single-qubit pure states~\cite{Reichardt04magic}.  In fact $\magic{\ket{\psi}}$ for all pure states $\ket{\psi}$ not stabilizer states, since all nonstabilizer single-qubit pure states $\ketbra{\psi}{\psi}$ have $\abs{x}+\abs{y}+\abs{z} > 1$.  Therefore all nonstabilizer pure states $\ket{\psi}$ give universality, $\magic{\ket{\psi}}$.

A natural question is whether the same type of reduction holds even for mixed states.  For brevity, call a state a nonstabilizer state if it is not a mixture of stabilizer states.  Given a nonstabilizer multi-qubit mixed state, can it be reduced to a single-qubit nonstabilizer mixed state, using postselected stabilizer operations?  

It turns out that the answer is no.  The multi-qubit question for mixed states does not reduce to the single-qubit question (\thmref{t:twoqubitcounterexample}).  We present examples of nonstabilizer mixed states for which every two-to-one-qubit stabilizer reduction gives a mixture of stabilizer states.  It is not known whether it is possible to achieve universality using these states in some more direct way.

Seven representative examples are given in \tabref{f:counterexamples}.  In this section, we will prove that \thmref{t:twoqubitcounterexample} holds for each of these states.  They were found by a geometrical argument considering the solid polyhedron $\mathcal O_n$ of convex combinations of all $n$-qubit stabilizer states (for $n=2$), a generalization of the solid octahedron $\mathcal O_1$ consisting of mixtures of single-qubit stabilizer states.

\begin{table}
\caption{Seven representative examples of states for which \thmref{t:twoqubitcounterexample} holds.  Each $\rho_i$ is $\tfrac{1}{4}II$ plus the fraction $f$ times the listed coordinates (tensor products implied).  For example, $\rho_1 = \tfrac{1}{4} II + \tfrac{1}{12} (IY + IZ - XX + YX + ZX)$.}\label{f:counterexamples}
\centering
\begin{tabular}{c | c @{\quad} c @{\quad} c @{\quad} c @{\quad} c @{\quad} c @{\quad} c @{\quad} c @{\quad} c @{\quad} c @{\quad} c @{\quad} c @{\quad} c @{\quad} c @{\quad} c}
\hline \hline 
$f$ & IX& IY& IZ& XI& XX& XY& XZ& YI& YX& YY& YZ& ZI& ZX& ZY& ZZ\\
\hline
1/12& 0 & 1 & 1 & 0 &-1 & 0 & 0 & 0 & 1 & 0 & 0 & 0 & 1 & 0 & 0 \\
1/76& 1 & 2 &-2 & 2 &-6 &-1 &-1 & 6 &-2 &-1 & 1 & 1 & 3 & 2 & 2 \\
1/72& 2 & 1 & 0 &-2 & 0 & 6 &-3 & 3 & 0 & 2 &-2 & 6 &-1 &-2 & 0 \\
1/76& 2 & 5 &-1 &-2 &-1 & 2 & 2 & 5 &-2 &-5 &-1 & 1 & 2 & 1 &-1 \\
1/52& 3 & 1 & 1 &-3 &-1 &-3 &-3 & 3 &-1 & 3 &-3 & 3 &-1 &-3 & 3 \\
1/60& 2 & 2 & 2 & 1 &-3 & 0 & 1 & 6 &-2 &-2 & 1 & 1 & 0 &-3 & 1 \\
1/52& 1 &-1 & 1 &-3 & 3 &-1 & 3 & 1 & 1 &-1 &-3 &-1 & 3 & 1 &-1 \\
\hline \hline
\end{tabular}
\end{table}

\subsection{Stabilizer reductions from $n$ qubits to one qubit}

To begin, we need to characterize the possible ways of using stabilizer operations to reduce an $n$-qubit state down to a single-qubit state.  We will show that, without loss of generality, one may use only Clifford group unitaries and postselected Pauli measurements.  It cannot be necessary to add ancilla qubits, or to adapt processing according to measurement results.  

\begin{lemma} \label{t:noancillasneeded}
For any postselected stabilizer procedure taking $\rho$ an $n$-qubit state to a one-qubit nonstabilizer state, there exists another procedure taking $\rho$ to a one-qubit nonstabilizer state that consists of an $n$-qubit Clifford unitary, followed by projecting the last $n-1$ qubits onto $\ket{0^{n-1}}$.
\end{lemma}

\begin{proof}
We first remark that without loss of generality all measurements in the reduction procedure are postselected.  Indeed, if some measurement $\mathcal{M}_i$ is not postselected, then the final outcome will be a mixture of the outcome conditioned on an $\mathcal{M}_i = +1$ measurement result and the outcome conditioned on $\mathcal{M}_i = -1$.  If the final outcome is not a mixture of stabilizer states, then at least one of these conditioned states must not have been.

To complete the proof, we must only enforce the assumption that the reduction procedure works on the $n$ qubits without requiring any extra ancillas.  For any procedure using ancillas, there is another procedure that uses no ancillas, and that has identical output on all $n$-qubit stabilizer states, except possibly with a higher success probability.  

This fact is a consequence of the Gottesman-Knill stabilizer algebra formalism~\cite{Gottesman97}.  In order to track the evolution of an arbitrary system under stabilizer operations, it suffices to keep track of the stabilizer group and logical $X$ and $Z$ values for each of the $n$ unfixed qubits.  

Assume that the procedure has the following form: 
\begin{enumerate}
\item In Phase~1, measure certain Pauli operators supported on the $n$ original (logical) qubits.  (Measuring a Pauli operator $P$ means applying the projection $\tfrac{1}{2}(I + P)$.)
\item Then prepare ancillas $\ket{0^m}$, and in Phase~2 measure Pauli operators acting possibly on all $n+m$ qubits.  
\item Finally, apply a Clifford unitary and trace out all but the first qubit.
\end{enumerate}
This form may be assumed since applying a unitary $U$ then measuring $P$, has the same effect as measuring $U^\dagger P U$, then applying $U$.  

Next we show how to either eliminate or move to Phase~1 any measurements in Phase~2.  Assume that Phase~1 has completed, leaving possibly an arbitrary state on the first $n$ qubits; set the stabilizer group $S$ to be all strings of $Z$ or $I$ supported on the ancilla qubits.  Consider the first measurement in Phase~2, of a Pauli operator $P$.  There are a few cases:
\begin{enumerate}
\item If $P$ or $-P$ is in $S$, then the measurement either has no effect or succeeds with probability zero.  Eliminate this measurement. 
\item If the operator $P$ being measured commutes with all the stabilizer group elements but is not in the stabilizer group ($P \in N(S) \setminus S$), then there exists $Q \in S$ such that $P Q$ is supported on the first $n$ qubits.  Measuring $P Q$ has the same effect as measuring $P$.  Remove the measurement of $P$ from Phase~2 and add a measurement of $P Q$ to Phase~1.  
\item If $P$ anticommutes with a stabilizer group element, then for some ancillary qubit $i$, $P_i \in \{ X, Y \}$.  Assume $P_i = X$; the other case is similar.   Then the Pauli $P X_i$ has the identity on the $i$th qubit.  Consider the unitary $U = \Lambda_i(P X_i)$ that applies $P X_i$ controlled by the $i$th qubit.  $U$ is a Clifford unitary such that $U P U = X_i$.  Therefore applying $U$, measuring $X_i$ and applying $U^\dagger$ to the result, is equivalent to measuring $P$.  However, applying $U$ has no effect, since the control-qubit is $\ket 0$.  Therefore, we can simply measure $X_i$, and delay application of $U^\dagger$ to the end of the protocol, by commuting it past any remaining measurements in Phase~2.  This measurement can be eliminated, since preparing $\ket 0$ and measuring $X$ is the same as simply preparing $H \ket{0} = \ket{+}$, except with lower success probability.
\end{enumerate}
Thus we may assume that there are no measurements in Phase~2.  

At the end of Phase~1, there remains at most one degree of freedom in the first $n$ qubits, and this degree of freedom can be isolated with Clifford unitaries involving only those qubits.  No extra ancillas are necessary.  

Moreover, the measurements in Phase~1 can be assumed to all commute with each other.  If not, and $P$ and $Q$ are two successive anticommuting measurements, then as above $P$ can be moved into just $Z_i$ for some $i$ and $Q$ can be assumed to be $X_i$.  Then measuring $Q$ was unnecessary, for one could have just applied $H_i$ with the same effect, and higher success probability.
\end{proof}

The fact that ancilla preparation is unnecessary has two useful consequences.  First, this leaves only a finite number of different stabilizer operations that can be applied to reduce an $n$-qubit state down to a single-qubit state.  In our proof we will consider up to symmetries an exhaustive list of all possible stabilizer operations reducing two qubits to one.  Second, it implies that a counterexample for $n=2$ also gives a counterexample for $n>2$; take the same state but adjoin $\ket{0^{n-2}}$.  If the original state is not a mixture of stabilizer states, then nor is the $n$-qubit state---any such mixture would need to be trivial on the last $n-2$ qubits.  If no algorithm using stabilizer operations reduces the two-qubit state to a nonstabilizer single-qubit state, then the same will be true for the $n$-qubit state because it can in particular be prepared from the two-qubit state with ancilla preparation.

\subsection{Polyhedron $O_n$ of mixtures of $n$-qubit stabilizer states} \label{s:twoqubitstatepolyhedron}

Any $n$-qubit density matrix $\rho$ can be written as a real combination of the $n$-qubit Pauli operators, with the coefficient of a Pauli $P$ given by $c_P(\rho) = \tfrac{1}{2^n} \tr (P \rho)$.  The coefficient of $I$ is fixed to $1/2^n$ since $\tr \rho = 1$, but the other $4^n-1$ coordinates can vary.  The state stabilized by the stabilizer group $\mathcal{S}$ has density matrix $\tfrac{1}{2^n} \sum_{S \in \mathcal{S}} S$, i.e., it has coordinates $1/2^n$ for $S \in \mathcal{S}$ and 0 elsewhere.  

\begin{lemma} \label{t:countinglemma}
The number of different $n$-qubit stabilizer states is
\beq
N = 2^n \times (2^n+1)(2^{n-1}+1)\cdots(2+1)
 \enspace .
\eeq
The number of different $n$-to-1-qubit stabilizer reductions, up to normalization and application of Cliffords to the output, is $\tfrac{1}{6}(2^n-1)N$.
\end{lemma}

\begin{proof}
The expression for $N$ is a simplification of 
\beq
2^n \frac{(4^n-1)(\tfrac{4^n}{2}-2)(\tfrac{4^n}{4}-4)\cdots(\tfrac{4^n}{2^{n-1}}-2^{n-1})}
{(2^n-1)(2^n-2)\cdots(2^n-2^{n-1})}
 \enspace .
\eeq
Here, the initial factor of $2^n$ is for the number of different syndromes given an unsigned set of stabilizers.  The numerator is the number of ways of picking (in order) $n$ nontrivial, independent, commuting generators.  Given a stabilizer group, the denominator is the number of ways of picking in order generators.

According to \lemref{t:noancillasneeded}, the $n$-to-1-qubit stabilizer reductions correspond to choosing a stabilizer group of size $2^{n-1}$, up to normalization and application of Cliffords to the final, logical degree of freedom, i.e., picking $n-1$ independent commuting generators.  Therefore the count of such reductions is
\beq
2^{n-1} \frac{(4^n-1)(\tfrac{4^n}{2}-2)(\tfrac{4^n}{4}-4)\cdots(\tfrac{4^n}{2^{n-2}}-2^{n-2})}
{(2^{n-1}-1)(2^{n-1}-2)\cdots(2^{n-1}-2^{n-2})}
 \enspace ,
\eeq
which simplifies to $\tfrac{1}{6}(2^n-1)N$.  The factor $\frac 1 6$ arises because choosing one nontrivial element of the stabilizer group to be, say, logical $+X$ overcounts by six.
\end{proof}

We are interested in convex combinations of stabilizer states.  For $n=1$, mixtures of the six stabilizer states form the closed, solid octahedron $\mathcal O_1$ in three dimensions, shown fitting within the Bloch sphere in \figref{f:blochsphere}.  For $n=2$, $\mathcal O_2$ has $60$ vertices in a $15$-dimensional space (\figref{f:twoqubitstabilizerstates}).  

\begin{figure}
\begin{equation*}
\begin{array}{c @{\qquad} c @{\qquad} c}
\Big\{\begin{array}{cc}
II & IX \\
XI & XX
\end{array}\Big\}
&
\Big\{\begin{array}{cc}
II & IX \\
YI & YX
\end{array}\Big\}
&
\Big\{\begin{array}{cc}
II & IX \\
ZI & ZX
\end{array}\Big\}
\\
& & \\ 
\Big\{\begin{array}{cc}
II & IY \\
XI & XY
\end{array}\Big\}
&
\Big\{\begin{array}{cc}
II & IY \\
YI & YY
\end{array}\Big\}
&
\Big\{\begin{array}{cc}
II & IY \\
ZI & ZY
\end{array}\Big\}
\\
& & \\ 
\Big\{\begin{array}{cc}
II & IZ \\
XI & XZ
\end{array}\Big\}
&
\Big\{\begin{array}{cc}
II & IZ \\
YI & YZ
\end{array}\Big\}
&
\Big\{\begin{array}{cc}
II & IZ \\
ZI & ZZ
\end{array}\Big\}
\\
& & \\ 
\Big\{\begin{array}{cc}
II & XX \\
YZ & ZY
\end{array}\Big\}
&
\Big\{\begin{array}{cc}
II & YY \\
XZ & ZX
\end{array}\Big\}
&
\Big\{\begin{array}{cc}
II & ZZ \\
XY & YX
\end{array}\Big\}
\\
& & \\ 
\Big\{\begin{array}{cc}
II & XY \\
YZ & -ZX
\end{array}\Big\}
&
\Big\{\begin{array}{cc}
II & XZ \\
YX & -ZY
\end{array}\Big\}
&
\Big\{\begin{array}{cc}
II & XX \\
YY & -ZZ
\end{array}\Big\}
\end{array}
\end{equation*}
\caption{
Stabilizer groups for $15$ two-qubit stabilizer states.  The $60$ total two-qubit stabilizer states can be enumerated by listing the four possible sign choices for each of these groups.  For example, by switching the sign of XX for the last case, we get also the state $\tfrac{1}{4}(II - XX + YY + ZZ)$.
} \label{f:twoqubitstabilizerstates}
\end{figure}

Notice that $n$-to-1-qubit stabilizer reductions are linear from the original $(4^n-1)$-dimensional space into the four-dimensional space with basis $I, X, Y, Z$---the coordinate for the identity $I$ is included because the trace of the output is not necessarily one without nonlinear renormalization.  Therefore, the set of $n$-qubit states mapped into $\mathcal O_1$ by a given stabilizer reduction is convex.

\subsection{Counterexamples for $n = 2$}

Specialize now to $n = 2$.  Let us show that none of the states in \tabref{f:counterexamples} are stabilizer states, and that for each of them every two-to-one-qubit stabilizer reduction gives a stabilizer state.  

Every face of $\mathcal O_2$ has at least $15$ vertices.  Each of the counterexamples in \tabref{f:counterexamples} comes from finding a face $F$ of $\mathcal O_2$ such that for every stabilizer reduction to a single qubit, $F$ is \emph{not} mapped to a face of the cone of $\mathcal O_1$.  That such faces of $\mathcal O_2$ can be found intuitively seems quite reasonable.  Indeed, the alternative would be that, for every face, all $15$ or more vertices are mapped to only four vertices in the output, three vertices of $\mathcal O_1$ plus possibly $0$.  For, if a fourth vertex of $\mathcal O_1$ is in the image, then necessarily two vertices must oppose one another and cancel out, so the image of $F$ is not a face.  Intuitively, this alternative possibility seems unlikely.  However, it can in fact occur for some faces of $\mathcal O_2$; see Eq.~\eqnref{e:firstineq} below.

\subsubsection{Counterexamples lie outside $\mathcal O_2$}

In order to prove that a certain state $\rho$ is nonstabilizer, we need to check that $\rho$ is indeed a valid density matrix, and also need to exhibit a separating hyperplane, i.e., an inequality satisfied by every stabilizer state but violated by $\rho$.  In \tabref{f:hyperplanes}, we list seven inequalities satisfied by each of the $60$ two-qubit stabilizer states, but violated by the respective states of \tabref{f:counterexamples}.  To verify this, compute inner products of the inequality with the state coordinates.  

\begin{table}
\caption{Inequalities defining seven faces of the two-qubit stabilizer polyhedron $\mathcal O_2$.  Each inequality is satisfied by all of the $60$ two-qubit stabilizer states, but the $j$th inequality is violated by the $j$th state of \tabref{f:counterexamples}.  To explain the notation, for example the first inequality is $\tfrac{1}{4} \tr \rho (-II + IY + IZ - XX + YX + ZX) \leq 0$.  One can check that substituting the first state from \tabref{f:counterexamples} gives $1/6$.}\label{f:hyperplanes}
\centering
\begin{tabular}{c @{\quad} c @{\quad} c @{\quad} c @{\quad} c @{\quad} c @{\quad} c @{\quad} c @{\quad} c @{\quad} c @{\quad} c @{\quad} c @{\quad} c @{\quad} c @{\quad} c @{\quad} c}
\hline \hline
II&IX&IY&IZ&XI&XX&XY&XZ&YI&YX&YY&YZ&ZI&ZX&ZY&ZZ\\
\hline
-1& 0& 1& 1& 0&-1& 0& 0& 0& 1& 0& 0& 0& 1& 0& 0 \\
-2& 1& 1&-1& 1&-2& 0& 0& 2&-1&-1& 1& 0& 1& 1& 1 \\
-2& 1& 1& 0&-1& 0& 2&-1& 1& 0& 0&-1& 2&-1&-1& 0 \\
-2& 1& 2& 0&-1& 0& 1& 1& 2&-1&-2& 0& 0& 1& 0& 0 \\
-2& 1& 0& 0&-1& 0&-1&-1& 1& 0& 1&-1& 1& 0&-1& 1 \\
-3& 2& 2& 1& 1&-2& 0& 1& 3&-2&-2&-1& 1& 0&-2& 1 \\
-4& 2&-1& 2&-3& 3&-2& 3& 1& 1&-2&-3&-1& 3& 2&-1 \\
\hline \hline
\end{tabular}
\end{table}

\subsubsection{Counterexamples reduce to mixtures of single-qubit stabilizer states} \label{s:counterexamplesreduce}

Next, we claim that for each counterexample from \tabref{f:counterexamples}, indeed every two-to-one-qubit stabilizer reduction outputs a mixture of stabilizer states.  This can be checked by enumeration, since by \lemref{t:noancillasneeded} there are only a finite number of stabilizer reductions that need to be considered.  In fact, there are exactly $30$ stabilizer reductions to check.  Each reduction corresponds to measuring one of the $15$ nontrivial two-qubit Pauli operators, postselecting on outcome $\pm1$.  This leaves one degree of freedom uniquely defined up to single-qubit stabilizer operations.  

Of course, these checks can easily be done on a computer, but it is worth understanding the algebra involved.  We will give a simple example that should elucidate the general situation.  

Consider, e.g., the state $\rho = \tfrac{1}{4}II + \tfrac{1}{12}(IY + IZ - XX + YX + ZX)$.  On measuring $ZZ$, postselecting on a $+1$ outcome, the unnormalized state becomes 
\begin{equation}\begin{split}
\tfrac{1}{2} (II + ZZ) \rho \, \tfrac{1}{2} (II + ZZ) 
&= \tfrac{1}{8}\big(II + ZZ + \tfrac{1}{3}(IZ + ZI - XX + YY + XY + YX)\big) \\
&= \tfrac{1}{4}\tfrac{1}{2}(II + ZZ)\big(II + \tfrac{1}{3}(IZ - XX + XY)\big)
 \enspace .
\end{split}\end{equation}
Applying a CNOT from the first qubit into the second in order to remove the fixed $ZZ$ stabilizer, the state becomes $\tfrac{1}{4}(I-\tfrac{1}{3}(X + Y + Z)) \otimes \tfrac{1}{2}(I + Z)$.  And indeed, the sum of the absolute values of the $X,Y,Z$ components of the first qubit equals the $I$ component, so the first qubit is a mixture of stabilizer states.  

We could have shortened the above calculation.  Notice that $IZ$, $ZI$, $XX$, $-YY$, $XY$, $YX$ are the nontrivial two-qubit Paulis commuting with $ZZ$, besides $ZZ$ itself.  Then compare $\rho$'s coordinates $c_{II} + c_{ZZ}$ to 
\beq \label{e:coordaverage}
\abs{c_{IZ}+c_{ZI}} + \abs{c_{XX}-c_{YY}} + \abs{c_{XY}+c_{YX}}
 \enspace .
\eeq
Both equal $1/4$.  

The general procedure follows immediately; we need to compare the sums of the absolute values of the appropriate averages of the coordinates of $\rho$, where the averages are over the Paulis commuting with the measured operator $P$ and differing by $P$.  For example, projecting $\rho$ by $\tfrac{1}{2}(II - YZ)$, we need to compare $c_{II} - c_{YZ}$ to 
\beq
\abs{c_{XX} - c_{ZY}} + \abs{c_{XY} + c_{ZX}} + \abs{c_{IZ} - c_{YI}}
 \enspace .
\eeq

One notices from these calculations an interesting property of $\rho$.  Let $S$ be the set of nontrivial Paulis $P$ such that $c_P(\rho) \neq 0$.  Then,
\begin{enumerate}
\item No two elements of $S$ commute.
\item The difference between any two elements of $S$ commutes with neither.  
\item Exactly three elements of $S$ commute with any nontrivial Pauli outside $S$.
\end{enumerate}
The first property implies that after measuring any element of $S$ and postselecting on either $+1$ or $-1$, the remaining degree of freedom is a fully mixed state---since the relevant sum of averaged coordinates is zero.  The second property, which is a consequence of the first property, implies that after measuring a Pauli $P$ not in $S$, it is impossible to have both $c_Q$ and $c_{PQ}$ nonzero for $Q$ commuting with $P$.  Therefore, the relevant comparison gives $\tfrac{1}{4} \geq \tfrac{1}{12} + \tfrac{1}{12} + \tfrac{1}{12}$, so the reduced state lies inside $\mathcal O_1$.  In fact, the third property implies that it lies on the boundary of $\mathcal O_1$.

Thus, these properties of $\rho$ suffice to prove that any two-to-one-qubit stabilizer reduction on $\rho$ outputs a mixture of stabilizer states.  We do not know of similarly concise proofs for the other states in \tabref{f:counterexamples}, but the calculation of course still reduces to summing averages of coordinates as in Eq.~\eqnref{e:coordaverage}.

\subsubsection{Structure of $\mathcal O_2$ by computer analysis}

The counterexamples from \tabref{f:counterexamples}
were found by using the {\tt cdd} software for polyhedral computations~\cite{Fukudacdd}.  On input the $60$ two-qubit stabilizer states, {\tt cdd} outputs the 22,320 external faces of their polyhedral convex hull.  Most of these faces are symmetrical under two-qubit Clifford operations; to determine a minimum set of representatives, we repeatedly chose a random two-qubit Clifford and reduced the faces modulo that symmetry.  After a small number of iterations, only eight faces remained.  Seven of these are those displayed in \tabref{f:hyperplanes}, and the eighth is given by
\beq \label{e:firstineq}
\tfrac{1}{4}\tr \rho(-II + IX + IY + IZ + XI - XX - XY - XZ) \leq 0
 \enspace .
\eeq
None of these remaining faces are symmetrical to each other, because two-qubit Cliffords can only permute their coordinates by conjugation, possibly also changing the sign $\pm 1$ of a coordinate.  Therefore, no two inequalities with differing II coordinates can be symmetrical to each other.  The inequality of Eq.~\eqnref{e:firstineq} and the first inequality of \tabref{f:hyperplanes} cannot be symmetrical to each other since they have a different number of nonzero coordinates.  No two of the inequalities in \tabref{f:hyperplanes} with II coordinate of $-2$ can be symmetrical since they either have different numbers of nonzero coordinates or have coordinates with different magnitudes.

Next we chose an element in the center of each face and pushed it out from $\tfrac{1}{4}II$ as far as possible such that every two-to-one stabilizer reduction still outputted only stabilizer mixtures.  This procedure worked for all the hyperplanes except that of Eq.~\eqnref{e:firstineq}.  For the state in the center of this face, there does exist a two-to-one reduction leading to a state lying on the boundary of $\mathcal O_1$ with equal $x,y,z$ coordinates.  Since we do not know the limits of distillation in this direction, it is unknown if applying this reduction to a state moved slightly outward leads to universality.

\subsection{Direct distillability of multi-qubit states}

We have \emph{not} shown that the states in \tabref{f:counterexamples} cannot be distilled to give universality, only that such a distillation procedure could not start by reducing from two qubits to one.  What more can be said about these examples, particularly the first one, having the most apparent structure?  

For more than two qubits, the calculation of whether or not the output lies inside $\mathcal O_1$ still simplifies to an equation like Eq.~\eqnref{e:coordaverage}, except within each of the three terms we average over a set of $2^{n-1}$ coefficients.  We have checked by exhaustive enumeration of all size-eight abelian subgroups of the Paulis on four qubits that for none of the states in \tabref{f:counterexamples} do two copies allow a reduction to a nonstabilizer state.  We have also verified this for several pairs of \emph{different} states from \tabref{f:counterexamples}.  

Unfortunately, this does not come close to proving that $\magic \rho$ is false for any of these states.  Currently there are essentially almost no nontrivial upper bounds on the power of magic states distillation even in the single-qubit case.  The only exception, to the author's knowledge, is recent work by Campbell and Browne~\cite{CampbellBrowne09magic}.  They show that for any fixed $x, y, z > 0$, and any fixed stabilizer code, there exists an $\epsilon > 0$ such that the distillation procedure based on this code fails for states $\rho(f x, f y, f z)$ when $\abs f \leq \frac{1+\epsilon}{x+y+z}$.  The possibility remains, though, that $\epsilon$ approaches $0$ as the code size increases.  

An interesting special case of the multi-qubit magic states distillation problem is the distillability of unentangled states.  Say that we are allowed to prepare copies of each of a set of states $\rho_1, \ldots, \rho_k$.  When do we have $\magic{\otimes_i \rho_i}$?  As a simple example, say we can prepare both $\rho(x,y,z)$ and $\rho(x,-y,z)$.  Reflection across the $xz$ plane is not a stabilizer operation, so these two states are generally not equivalent under stabilizer operations.  Now letting $\Pi_{\mathrm{even}} = \ketbra{00}{00} + \ketbra{11}{11}$,
\begin{equation}
\begin{split}
\Pi_{\mathrm{even}} (\rho(x,y,z) \otimes \rho(x,-y,z)) \Pi_{\mathrm{even}}
&= \tfrac{1}{4} \left(\begin{smallmatrix}
(1+z)^2 & x^2+y^2 \\
x^2+y^2       & (1-z)^2
\end{smallmatrix}\right) \\
&\propto \rho(\tfrac{r^2}{1+z^2},0,\tfrac{2z}{1+z^2}) \enspace ,
\end{split}
\end{equation}
where $r^2 = x^2 + y^2$.  Thus immediately the $y$ coordinate is zeroed out, and we essentially have Eq.~\eqnref{e:yzero} with $r$ in place of $x$.  Thus directly $\magic{\rho(x,y,z) \otimes \rho(x,-y,z)}$ provided $\max\{\sqrt{x^2+y^2}+z, x+\sqrt{y^2+z^2}\} > 1$.

\section{Distillation of unknown states}

Note that the success of the method for proving \thmref{t:rotations} depends strongly on us knowing the state $\rho$ \emph{exactly}, and that precisely the same state can be prepared repeatedly.  Any small errors will be amplified quickly in the angle-doubling step.  These assumptions are very strong, and are probably not physically justified.  For practical applications, the definition of $\magic \rho$ should be revisited to incorporate other conditions, including stability.  The exact conditions may depend on the application.  For example, in the threshold proof for concatenated distance-three codes in Ref.~\cite{Reichardt05distancethree}, dealing with the possibility of asymmetrically erroneous states requires: 

\begin{theorem} \label{t:practicalmagicrandom}
For any constant $\delta > 0$, perfect stabilizer operations with adaptive classical control, together with the ability to prepare states $\rho_i = \rho(f_i/\sqrt 3, f_i/\sqrt 3, f_i/\sqrt 3) = \tfrac{1}{2}\big(I + \tfrac{f_i}{\sqrt{3}}(X+Y+Z)\big)$ with each $f_i$ unknown but at least $(1+\delta) \sqrt{3/7}$, gives quantum universality.  
\end{theorem}

Thus it is enough to have lower bounds on the fidelities of the prepared states with $\ket T$, and the states need not be identical.  Moreover, in this case, the bound on the allowed error rate turns out to be the same for these nonidentical states as for the identical states assumed in \thmref{t:BravyiKitaev04}.  

Note that some assumptions that are impractical at the physical level become relevant at the encoded level, when we try to apply magic states distillation on top of a stabilizer operation fault-tolerance scheme.  An example is the assumption of perfect stabilizer operations.  See \secref{s:lowerbounds}.  Additionally, it might not be surprising that the distillation model becomes more delicate as the limits of distillation are approached.  Even delicate, artificial models can be of interest when we use magic states distillation to consider noise threshold upper bounds, in \secref{s:upperbounds}.  Still, it is possible that the current techniques are unnecessarily fragile, and that there exist more direct, practical and efficient methods.

\begin{proof}[Proof of \thmref{t:practicalmagicrandom}]

\thmref{t:practicalmagicrandom} is an extension of \thmref{t:BravyiKitaev04}, from~\cite{BravyiKitaev04}.  It is based on the same five-qubit code distillation scheme discussed in \secref{s:fivequbit}.  
Denote by $[f]^s$ the symmetric sum of $s$-tuples of the variables $f_1, \ldots, f_5$, i.e., 
\beq
[f]^s \equiv \sum_{\substack{S \subseteq \{1,2,3,4,5\} \\ \abs{S} = s}} \prod_{i \in S} f_i
 \enspace .
\eeq

Take five prepared states, $\otimes_{i=1}^{5} {\rho_i}$, and use stabilizer operations to project into the codespace of the five-qubit code, then decode the logical qubit.  A simple calculation gives that the probability of success is 
\beq
p_\text{success} = \tfrac{1}{48}\big(3 + [f]^4\big)
 \enspace .
\eeq
The $x$, $y$, $z$ coordinates of the output state, conditioned on success, are the same, equal to 
\beq
\frac{1}{\sqrt{3}} \cdot \frac{1}{p_\text{success}} \frac{-1}{48}\big([f]^3 - 2 [f]^5\big)
 \enspace .
\eeq
These coordinates are negative, but can be rotated back to the positive octant with a stabilizer operation.  Then the output state is $\rho(f_{out}/\sqrt 3, f_{out}/\sqrt 3, f_{out}/\sqrt 3)$, where 
\beq
f_\text{out} = \frac{[f]^3 - 2[f]^5}{3+[f]^4}
 \enspace .
\eeq

Now note that $p_\text{success}$ is monotonely increasing in each $f_i$, so distillation remains efficient when the $f_i$ are unequal.  

Also, simple algebra gives that $\partial f_\text{out} / \partial f_i > 0$, so improving any of the input fidelities can only improve the output fidelity.  
Indeed, differentiate $f_ \text{out}$ with respect to $f_5$ -- other derivatives are related by symmetry.  Use the quotient rule $d (a/b) = \tfrac{1}{b^2}(b\, da - a\, db)$.  The numerator, which does not involve $f_5$, is, after simplifications, 
\beq
f_1 f_2 \left( 3 - f_1 f_2 f_3 f_4 - f_3 f_4 - \tfrac{1}{3} f_1 f_2 f_3^2 f_4^2 - \tfrac{1}{3} f_1 f_2(f_3^2 + f_4^2) \right) + \textrm{symmetrical terms}
 \enspace .
\eeq
Each term is nonnegative when the $f_i \in [0,1]$, implying that $f_\text{out}$ is monotone in each $f_i$. 
\end{proof}

The assumption of \thmref{t:practicalmagicrandom}, that each prepared $\rho_i$ has equal $x$, $y$, $z$ coordinates, can be guaranteed by randomly applying $I$, $T$, or $T^2$ each with probability $1/3$ independently to each prepared $\rho_i$.  However, the ability to apply perfect Clifford unitaries may \emph{not} imply the ability to apply a \emph{random} perfect Clifford unitary.  Depending on the application, this symmetrization may not be innocuous.  

For example, one important class of schemes for achieving fault tolerance is based on postselection~\cite{Knill05, Reichardt06thesis, AliferisGottesmanPreskill07postselected, Reichardt07algorithmica}.  In such schemes one can apply stabilizer operations to encoded qubits.  These operations can be made arbitrarily accurate, but only \emph{conditioned} on some heralded random event.  That is, after trying to applying to apply an operation, the experimenter receives a message whether or not the operation succeeded.  The details are not important here.  The problem, though, is that the success probability can depend on the operation being applied.  Thus, if we try to apply $I$, $T$, or $T^2$ each with probability $1/3$, there is no guarantee that the probabilities conditioned on success will also be each $1/3$.  

Now the fact that there is a working scheme for distilling $\rho_1 \otimes \cdots \otimes \rho_n$ that starts by applying $I$, $T$ or $T^2$ at random to each qubit implies that there exists some \emph{fixed} sequence of unitaries $U_1 \otimes \cdots \otimes U_n$ that works at least as well.  If we knew this sequence, we would have no need to randomize.  Unfortunately, this sequence a priori could depend on the states $\rho_i$ in an arbitrary manner.  Without knowing the states exactly, therefore, we cannot derandomize the distillation scheme.  In fact, though, Ref.~\cite{Reichardt07algorithmica} shows that the $\ket T$-state distillation scheme \emph{can} be derandomized, albeit at a cost: 

\begin{theorem}[{\cite[Theorem~1]{Reichardt07algorithmica}}] \label{t:practicalmagicdeterministic}
There exists a constant $\epsilon > 0$ such that perfect stabilizer operations with adaptive classical control, together with the ability to prepare (unknown) states $\rho_i$ each with fidelity $\geq 1-\epsilon$ with $\ket T$, gives quantum universality.  

More explicitly, $\rho_i = \rho(x_i, y_i, z_i)$, a sufficient condition for quantum universality is that for each~$i$, 
\beq
\max_i \max \big\{ \abs{\tfrac{1}{\sqrt{3}} - x_i}, \abs{\tfrac{1}{\sqrt{3}} - y_i}, \abs{\tfrac{1}{\sqrt{3}} - z_i} \big\} \leq 0.0527
 \enspace .
\eeq
\end{theorem}

In fact, by considering a decoding circuit for the five-qubit code, Ref.~\cite{Reichardt07algorithmica} shows that not all stabilizer operations are necessary for simulating universal quantum computation.  In \thmref{t:practicalmagicdeterministic}, it suffices to have the operations CNOT, Hadamard, preparation of $\ket 0$ and measurement in the $\ket 0$, $\ket 1$ basis.  

In the next section, we will summarize applications of magic states distillation to quantum fault-tolerance schemes.  For lower bounds on the fault-tolerance threshold, statements like \thmref{t:practicalmagicrandom} or \thmref{t:practicalmagicdeterministic} are typically needed.  For placing limits on \emph{upper} bounds, though, i.e., on attempts to prove the impossibility of reliable quantum computation at high noise rates, theorems such as \thmref{t:BravyiKitaev04}, \thmref{t:rotations} and \thmref{t:fivequbittwisted}, which assume the state $\rho$ is known exactly and can be prepared repeatedly, are still of use.

\section{Fault-tolerance applications} \label{s:faulttolerance}

The main application of magic states distillation is to fault-tolerant quantum computing, or reliable computing in the presence of noisy gates.  In particular, it clearly addresses the problem of achieving universality using noisy ancilla preparation.  It does assume perfect stabilizer operations, though, which is certainly not realistic.  This assumption can be justified in two different contexts: 

1. Certain arguments for upper-bounding the tolerable noise rate (or ``threshold") assume that stabilizer operations are perfect and only the extra operation required for universality is noisy.  This is optimistic, but sufficient for an upper bound.  \secref{s:upperbounds} below considers the relationship of magic states distillation to these arguments.  

2. Assumptions, like perfect stabilizer operations, which are unrealistic at the physical level can sometimes be justified at higher levels of encoding in a fault-tolerant concatenated coding scheme.  In particular, it is possible that there are two different noise thresholds, one threshold for reliable stabilizer operations and a separate threshold for reliable universal quantum computation.  If the physical noise is below the threshold for reliable stabilizer operations, then we can assume that stabilizer operations are in fact perfect---not at the physical level, but at some level of encoding.

\subsection{Fault-tolerance threshold lower bounds and estimates} \label{s:lowerbounds}

Assuming perfect stabilizer operations, magic states distillation gives universality by using noisy, single-qubit  ancilla states.  However, these noisy ancillas need to be prepared not at the physical level, but at the same higher level of encoding at which the logical stabilizer operations are reliable.  

\begin{figure}
\begin{center}
\includegraphics[scale=.4]{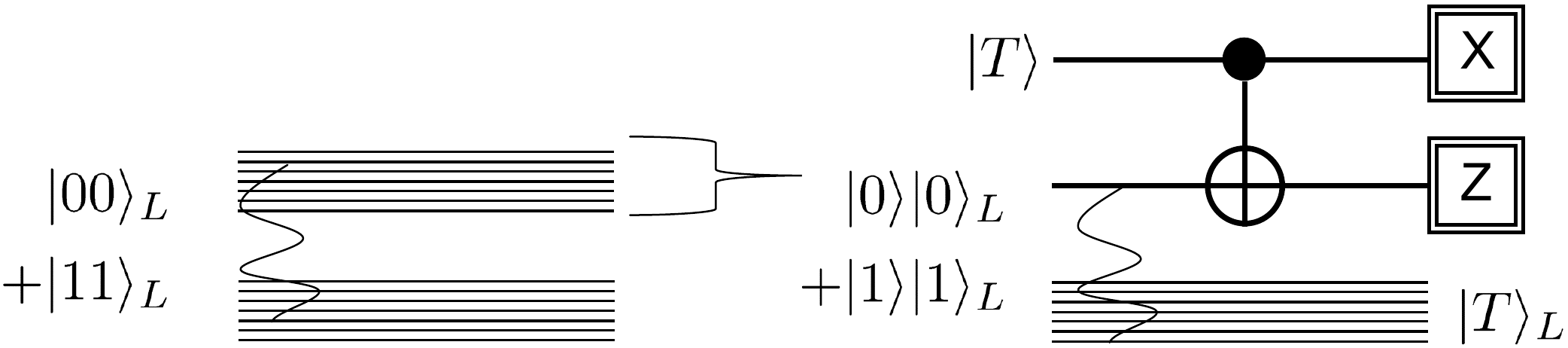}
\end{center}
\caption{Knill's method for achieving universality by teleporting into an encoding.  If the X and Z measurements are not postselected, then a logical correction (not shown) might be required.}
\label{f:universalityreduction}
\end{figure}

How can one reliably encode noisy ancillas?  Following Knill~\cite{Knill04schemes,Knill05}, we use perfect encoded stabilizer operations to create an encoded Bell pair $\tfrac{1}{\sqrt{2}}(\ket{00}_L + \ket{11}_L)$ (the subscript $L$ denoting ``logical").  Then we decode one half from the bottom up, ideally obtaining $\tfrac{1}{\sqrt{2}}(\ket{0}\ket{0}_L + \ket{1}\ket{1}_L)$.  Then prepare a qubit in a ``magic" state like $\ket H$ or $\ket T$ and teleport it into the encoding, using a physical CNOT gate and two single-qubit measurements.  See \figref{f:universalityreduction}.  If there is no noise, then the output state will be $\ket{H}_L$ or $\ket{T}_L$.  At that point, both stabilizer operations and ancilla preparation can be done at an encoded level, so encoded universality follows.  

In the presence of noise, the noise too will be teleported into the encoding, i.e., into logical noise.  As long as it is not too high, it can be distilled away at the encoded level, using (perfect) encoded stabilizer operations.  This noise can come from three places: noise in the prepared single-qubit ancilla, noise in the physical teleportation circuit, and noise in the decoded half of the Bell pair.  As long as the total noise from these sources is not too large, magic states distillation will succeed.  Fortunately, magic states distillation can tolerate very high amounts of noise---for example, $\tfrac{1}{2}(1-\sqrt{3/7}) \approx 17.2\%$ depolarizing noise on $\ket T$, by \thmref{t:BravyiKitaev04}.  Therefore, it often turns out that the threshold for universal quantum computation is the same as that for just stabilizer operations.  The bottleneck is in achieving reliable stabilizer operations.  

This is a useful technique because it is easier to prove a noise threshold for stabilizer operations alone than for a full universal scheme.  For fault-tolerance schemes using quantum stabilizer codes, only physical stabilizer operations are required to achieve reliable encoded stabilizer operations.  Stabilizer operations are easy to work with because Pauli errors propagate through them linearly.  Moreover, these operations can be simulated efficiently, by the Gottesman-Knill theorem---meaning that we can run classical simulations to \emph{estimate} the fault-tolerance threshold for stabilizer operations.  Steane and Knill, among many others, have run extensive simulations of this type to determine threshold estimates~\cite{Steane03, Knill05}.  

Threshold proofs and estimates simplify using this reduction because magic states distillation lets us skip the fault-tolerance hierarchy for universal quantum computing operations.  That is, with a concatenated-coding fault-tolerance scheme, stabilizer operations at $k$ levels of encoding are implemented in terms of stabilizer operations at $k-1$ levels of encoding.  Typically, the additional operation needed to obtain universality is also implemented at level $k$ in terms of the same operation and stabilizer operations at level $k-1$.  Using magic states distillation, though, we can instead obtain universality at level $k$ using only level-$k$ stabilizer operations and some sort of universality operation at level $0$, in this case, certain noisy ancilla preparations.  

This reduction, from universal fault tolerance to stabilizer operation fault tolerance, does not necessarily always work, because it requires that we be able to decode one half of an encoded Bell pair without introducing too much noise.  It is possible that we can prepare a perfect encoded Bell pair but cannot decode half of it without losing control of the noise.  In the schemes that have been studied~\cite{Knill05,Reichardt05distancethree}, however, this has not seemed to happen.  The decoding operation was straightforward to analyze rigorously in Ref.~\cite{Reichardt05distancethree} because decoding blocks independently cannot create \emph{correlated} errors, which are the main obstacle to proving a threshold for stabilizer operations.  

There are different tricks that might be useful, too, in decoding a state; for example, instead of correcting any detected errors, one might postselect on no detected errors.  If any errors are detected, then throw the Bell pair away and start over.  This can adversely impact the overhead if applied injudiciously, but it might be a reasonable technique to apply in a limited fashion.  For example, in a recursive decoding scheme, one might only postselect on no detected errors in decoding the last few levels.

Some technical concerns arise in applying magic states distillation to fault-tolerance.  For example, besides knowing $\magic{\rho}$, we are also interested in the stability of the procedure, either because we cannot repeatedly prepare exact copies of $\rho$ or perhaps because our knowledge of $\rho$ has limited accuracy~\cite{Reichardt05distancethree}.  These concerns are addressed by \thmref{t:practicalmagicrandom} and \thmref{t:practicalmagicdeterministic}.  

Finally, note that while fault-tolerance schemes often use concatenated coding, and magic states distillation can also be phrased as projection into a certain code space, the two codes need bear no relationship to each other.

\subsection{Fault-tolerance threshold upper bounds} \label{s:upperbounds}

Giving upper bounds for the fault-tolerance threshold, with a given set of operations and a given noise model, is difficult.  There have been only a few approaches, and these tend to be tied delicately to a particular model.  For example, Aharonov et al.~show that a useful noisy quantum circuit can only have logarithmic depth if fresh ancillas are not allowed to be introduced during the computation~\cite{AharonovBenOr96,AharonovBenorImpagliazzoNisan96}.  In practical quantum computing schemes, though, it is possible to initialize ancillas during the computation.  Razborov shows that the tolerable noise rate, of circuits with more than logarithmic depth, can be at most $1/2$ for a gate set with gates of fan-in two~\cite{Razborov}.  This approach does not allow for noiseless classical control based on measurement results, though, which is common in proposed experimental implementations of quantum computers.  Moreover, interesting problems, including factoring, \emph{can} be solved with logarithmic-depth quantum circuits, aided by classical computation~\cite{CleveWatrous00}.  

Harrow and Nielsen~\cite{HarrowNielsen} ask how much depolarizing noise can be tolerated by a two-qubit gate before it loses its power to generate entanglement; they find that the CNOT is the most resilient two-qubit gate, but does not tolerate independent depolarizing noise higher than 74\%. (Virmani, Huelga and Plenio improve this to $2/3$ with a more careful entanglement requirement~\cite{VirmaniHuelgaPlenio}.)  Against simultaneous depolarizing noise, they find that the threshold is at most $8/9$, or $1/2$ for a somewhat-adversarial noise model, an optimal noise process including correlated two-qubit noise.

Virmani et al.~\cite{VirmaniHuelgaPlenio} assume that stabilizer operations, including the CNOT gate, are perfect, and ask how much noise can be tolerated in an additional gate used to achieve universality.  They show that the $\pi/8$ gate, $\exp(i \tfrac{\pi}{8}Z)$, with $(\sqrt{2}-1)/2\sqrt{2} \approx 14.6\%$ or more worst-case noise, or twice that amount of dephasing noise, becomes a convex combination of stabilizer operations and so this gate set can be simulated classically.  Among all the rotations $\exp(i \tfrac{\theta}{2}Z)$, the $\pi/8$ gate is the most resistant to dephasing noise according to their criterion.  The advantage of this approach, and also that of Harrow and Nielsen, is that it easily allows for the incorporation of noiseless classical control into the model.  

Buhrman et al.~extend these results to a depolarizing noise channel~\cite{BuhrmanCleveLaurentLindenSchrijverUnger06}.  Again, assume that stabilizer operations are perfect, and assume that a noisy single-qubit gate is used to achieve universality.  They show that the $\pi/8$ gate with $(6 - 2 \sqrt 2)/7 \approx 45.3\%$ or more depolarizing noise becomes a convex combination of stabilizer operations.  And again, the $\pi/8$ gate is the most noise-resistant single-qubit gate.  Therefore, $45.3\%$ is an upper bound on the noise threshold in this model.  

Magic states distillation (\thmref{t:pipereighttheorem}) shows the limit of the techniques of~\cite{VirmaniHuelgaPlenio, BuhrmanCleveLaurentLindenSchrijverUnger06}.  Both their upper bounds are tight; with any less noise one gets universal quantum computation.  Since from \secref{s:lowerbounds} we typically expect the threshold bottleneck to be in achieving perfect stabilizer operations, this may not be very surprising.  

\begin{proof}[Proof of \thmref{t:pipereighttheorem}]
The upper bounds are due to~\cite{VirmaniHuelgaPlenio, BuhrmanCleveLaurentLindenSchrijverUnger06}.  

The $\pi/8$ gate with less than $(\sqrt{2}-1)/2\sqrt{2}$ worst-case probabilistic noise, or twice that amount of dephasing noise, takes $\ket{+}$ to a state $\rho(x,x,0)$ with $x > 1/2$, implying universality together with perfect stabilizer operations by \thmref{t:Reichardt04magic}.  

The $\pi/8$ gate with $45\%$ depolarizing noise, however, takes $\ket{+}$ to a state well inside the octahedron $\mathcal O_1$ of \figref{f:blochsphere}.  Instead, inspired by the Jamiolkowski isomorphism, apply the noisy gate to the second half of a Bell pair, which is a stabilizer state.  If the depolarizing noise rate is less than $(6 - 2 \sqrt 2)/7$, then the output two-qubit state lies outside of $\mathcal O_2$ (defined in \secref{s:twoqubitstatepolyhedron}).  Moreover, there does exist a two-to-one-qubit stabilizer reduction giving a state outside $\mathcal O_1$; simply apply the parity-check procedure of \secref{s:universality}.  Indeed, the renormalized output state at a depolarizing noise rate of $(6 - 2 \sqrt 2)/7 - \epsilon$ is computed to have $x, y, z$ coordinates of 
\beq
\tfrac{1}{10 + 6 \sqrt 2 + 21 \epsilon} \left( (1 + 2 \sqrt 2)(2 + 7 \epsilon), -2 \sqrt 2 (1 + 2 \sqrt 2 + 7 \epsilon), 0 \right)
 \enspace ,
\eeq
for which $\abs x +\abs y > 0$ when $\epsilon > 0$.  By \thmref{t:Reichardt04magic}, this state gives universality.  
\end{proof}

There are a number of questions still to answer about these threshold upper bound results.  We will just list a few of them:

\begin{enumerate}
\item
Assume $\mathcal E$ is a partly depolarized single-qubit unitary $\mathcal E(\rho) = (1-p) U \rho U^\dagger + p\tfrac{I}{2}\tr\rho$ (or, more generally, a mixture of unitaries $\mathcal E(\rho) = \sum_i p_i U_i \rho U_i^\dagger$), but is not a mixture of Cliffords.  Does $\mathcal E$ together with stabilizer operations suffice for achieving universality?  Certainly simply applying $\mathcal E$ to a single-qubit Pauli eigenstate may not suffice, for it could create, e.g., $\rho(x,y,z)$ with $x=y=z \in (\tfrac{1}{3}, \tfrac{1}{\sqrt 7}]$ which we do not know not how to distill.  None of the counterexamples from \secref{s:twoqubitstate} can be written as a single-qubit unitary on a stabilizer state followed by depolarizing noise.\footnote{Indeed, each counterexample has a nonzero coordinate for one of IX, IY, IZ, XI, YI or ZI, and it is simple to prove that the same is true after applying any two-qubit Clifford---the interesting case is the first example of \tabref{f:counterexamples}.}  Note that the last six elements in \figref{f:twoqubitstabilizerstates} correspond to the 24 single-qubit Clifford unitaries under the Jamiolkowski isomorphism, up to sign/syndrome choices.

For a noisy gate $\mathcal E$, stabilizer operations with adaptive classical control and $\mathcal E$ gives universality if and only if $\magic{({\boldsymbol 1} \otimes \mathcal E)(\ketbra \Psi \Psi)}$ for $\ket{\Psi} = \tfrac{1}{\sqrt 2}(\ket{00}+\ket{11})$.  Indeed, the ``if" direction is by definition of $\magic{\rho}$.  Only if: $\mathcal E$ being universal with stabilizer operations implies that we can efficiently approximate $\ket T$ to arbitrary accuracy, i.e., using $\poly(\log(1/\delta))$ gates to obtain precision $\delta$.  In particular, we can get within a constant of $\ket T$ using only a constant number of applications of $\mathcal E$.  Now assume we have only stabilizer operations and repeated preparation of $\rho = ({\boldsymbol 1} \otimes \mathcal E)(\ketbra{\Psi}{\Psi})$.  In our approximate preparation of $\ket T$, replace every application of $\mathcal E$ by teleportation into $\rho$, conditioning each measurement outcome so no correction is required.  Since the success probability of each teleportation is $1/4$ and only a constant number of teleportations are required, the expected overhead is a constant.  Once we have an approximation of $\ket T$, we obtain universality by distilling it, e.g., using \thmref{t:practicalmagicrandom}.  

In fact, if $\mathcal E$ is the $\pi/8$ gate with depolarizing noise, this equivalence is much simpler; teleporting into $({\boldsymbol 1} \otimes \mathcal E)(\ketbra{\Psi}{\Psi})$ can be accomplished deterministically.  Since the $\pi/8$ gate is in $C_3$, the set of operators that conjugate Paulis to Cliffords, any required correction is always a Clifford and can be applied~\cite{GottesmanChuang99teleportation}.  The depolarizing noise commutes past the correction.

Therefore, without loss of generality we may assume that $\mathcal E$ is applied to the second half of $\ket{\Psi}$.  But can one always distill the resulting state by first applying a two-to-one-qubit stabilizer reduction?

\item
If we do not assume perfect CNOT gates, then can we reduce the amount of error allowed on the single-qubit gate, such as a $\pi/8$ gate?  This is certainly sometimes the case if our gate set consists of preparation of noisy ancillas.  For example, if the CNOT model is bitwise independent depolarizing channels prior to the CNOT, then we of course will not achieve universality if the total depolarizing noise on the ancilla moves it into the Bloch sphere.  This criterion is probably not tight, however.  Can we get similar results for more interesting noise models, or for the $\pi/8$ gate?  Plenio and Virmani have recently studied this problem under the assumption that the fault-tolerance scheme uses the magic state ancillas in a certain way~\cite{PlenioVirmani08magic}.  They obtain noise threshold upper bounds, subject to this assumption, that are remarkably close to estimated threshold lower bounds.  
\end{enumerate}

\section{Conclusion}

Magic states distillation is important both because it describes the power of stabilizer operations, in terms of what more is necessary for achieving full quantum universality, and because of its application to quantum fault tolerance.  We have given improved magic states distillation procedures, reducing the set of single-qubit mixed states $\rho$ for which $\magic{\rho}$ is unknown.  We have also introduced the multi-qubit magic states distillation problem, and proved that it does not reduce to the single-qubit case.  We have used magic states distillation to prove that two noise threshold upper bounds are in fact tight.  

There remain many open problems, including most of those described in earlier papers~\cite{BravyiKitaev04,Reichardt04magic}.  Are there two-qubit nonstabilizer states that cannot be distilled to a single-qubit nonstabilizer state?  We have given some candidate states, one of which might have sufficient structure for an analysis.  In particular, it is interesting to specialize this question to those two-qubit states arising from the Jamiolkowski isomorphism.

\smallskip
Research conducted while the author was at the University of California, Berkeley, supported in part by NSF ITR Grant CCR-0121555, and ARO Grant DAAD 19-03-1-0082.

\bibliographystyle{alpha-eprint}
\bibliography{tun}

\newcommand{\etalchar}[1]{$^{#1}$}
\begin{thebibliography}{ABOIN96}
\expandafter\ifx\csname urlprefix\endcsname\relax\def\urlprefix{URL }\fi
\providecommand{\arxiv}[2][]{\href{http://arxiv.org/pdf/#2}{\texttt{arXiv:#2}}}
\providecommand{\doi}[2][]{\href{http://dx.doi.org/#2}{\texttt{doi:#2}}}

\bibitem[ABO96]{AharonovBenOr96}
Dorit Aharonov and Michael Ben-Or.
\newblock Polynomial simulations of decohered quantum computers.
\newblock In {\em Proc. 37th IEEE Foundations of Computer Science (FOCS)},
  pages 46--55, 1996.

\bibitem[ABOIN96]{AharonovBenorImpagliazzoNisan96}
Dorit Aharonov, Michael Ben-Or, Russell Impagliazzo, and Noam Nisan.
\newblock Limitations of noisy reversible computation, 1996,
  \href{http://www.arxiv.org/abs/quant-ph/9611028}{{\tt
  arXiv:quant-ph/9611028}}.

\bibitem[AG04]{AaronsonGottesman04}
Scott Aaronson and Daniel Gottesman.
\newblock Improved simulation of stabilizer circuits.
\newblock {\em Phys. Rev. A}, 70:052328, 2004,
  \href{http://www.arxiv.org/abs/quant-ph/0406196}{{\tt
  arXiv:quant-ph/0406196}}.

\bibitem[AGP08]{AliferisGottesmanPreskill07postselected}
Panos Aliferis, Daniel Gottesman, and John Preskill.
\newblock Accuracy threshold for postselected quantum computation.
\newblock {\em Quant. Inf. Comput.}, 8:181--244, 2008,
  \href{http://www.arxiv.org/abs/quant-ph/0703264}{{\tt
  arXiv:quant-ph/0703264}}.

\bibitem[BBC{\etalchar{+}}95]{BarencoBennettCleveDiVincenzoMargolusShorSleator%
SmolinWeinfurter95}
Adriano Barenco, Charles~H. Bennett, Richard Cleve, David~P. DiVincenzo, Norman
  Margolus, Peter~W. Shor, Tycho Sleator, John~A. Smolin, and Harald
  Weinfurter.
\newblock Elementary gates for quantum computation.
\newblock {\em Phys. Rev. A}, 52(5):3457--3467, 1995.

\bibitem[BCL{\etalchar{+}}06]{BuhrmanCleveLaurentLindenSchrijverUnger06}
Harry Buhrman, Richard Cleve, Monique Laurent, Noah Linden, Alexander
  Schrijver, and Falk Unger.
\newblock New limits on fault-tolerant quantum computation.
\newblock In {\em Proc. 47th IEEE Foundations of Computer Science (FOCS)},
  2006.

\bibitem[BK05]{BravyiKitaev04}
Sergey Bravyi and Alexei Kitaev.
\newblock Universal quantum computation with ideal clifford gates and noisy
  ancillas.
\newblock {\em Phys. Rev. A}, 71:022316, 2005,
  \href{http://www.arxiv.org/abs/quant-ph/0403025}{{\tt
  arXiv:quant-ph/0403025}}.

\bibitem[BMP{\etalchar{+}}00]{BoykinMorPulverRoychowdhuryVatan00}
P.~Oscar Boykin, Tal Mor, Matthew Pulver, Vwani Roychowdhury, and Farrokh
  Vatan.
\newblock A new universal and fault-tolerant quantum basis.
\newblock {\em Information Processing Letters}, 75:101--107, 2000.

\bibitem[CB09]{CampbellBrowne09magic}
Earl~T. Campbell and Daniel~E. Browne.
\newblock Neither magical nor classical?
\newblock unpublished, 2009.

\bibitem[CW00]{CleveWatrous00}
Richard Cleve and John Watrous.
\newblock Fast parallel circuits for the quantum fourier transform, 2000,
  \href{http://www.arxiv.org/abs/quant-ph/0006004}{{\tt
  arXiv:quant-ph/0006004}}.

\bibitem[Den01]{Dennis01distill}
Eric Dennis.
\newblock Toward fault-tolerant quantum computation without concatenation.
\newblock {\em Phys. Rev. A}, 63:052314, 2001,
  \href{http://www.arxiv.org/abs/quant-ph/9905027}{{\tt
  arXiv:quant-ph/9905027}}.

\bibitem[FLMR04]{fernandez-2004-2}
Jose~M Fernandez, Seth Lloyd, Tal Mor, and Vwani Roychowdhury.
\newblock Algorithmic cooling of spins: A practicable method for increasing
  polarization.
\newblock {\em International Journal of Quantum Information}, 2:461, 2004.

\bibitem[Fuk]{Fukudacdd}
Komei Fukuda.
\newblock cdd software.
\newblock Available from
  http://www.cs.mcgill.ca/{$\sim$}fukuda/soft/cdd\_home/cdd.html.

\bibitem[GC99]{GottesmanChuang99teleportation}
Daniel Gottesman and Isaac~L. Chuang.
\newblock Quantum teleportation is a universal computational primitive.
\newblock {\em Nature}, 402:390--393, 1999,
  \href{http://www.arxiv.org/abs/quant-ph/9908010}{{\tt
  arXiv:quant-ph/9908010}}.

\bibitem[Got98]{Gottesman97}
Daniel Gottesman.
\newblock A theory of fault-tolerant quantum computation.
\newblock {\em Phys. Rev. A}, 57:127, 1998,
  \href{http://www.arxiv.org/abs/quant-ph/9702029}{{\tt
  arXiv:quant-ph/9702029}}.

\bibitem[HN03]{HarrowNielsen}
Aram~W. Harrow and Michael~A. Nielsen.
\newblock How robust is a quantum gate in the presence of noise?
\newblock {\em Phys. Rev. A}, 68:012308, 2003,
  \href{http://www.arxiv.org/abs/quant-ph/0301108}{{\tt
  arXiv:quant-ph/0301108}}.

\bibitem[KLZ96]{KnillLaflammeZurek96}
Emanuel Knill, Raymond Laflamme, and Wojciech Zurek.
\newblock Accuracy threshold for quantum computation, 1996,
  \href{http://www.arxiv.org/abs/quant-ph/9610011}{{\tt
  arXiv:quant-ph/9610011}}.

\bibitem[KLZ98]{KnillLaflammeZurekProcRSocLondA98}
Emanuel Knill, Raymond Laflamme, and Wojciech Zurek.
\newblock Resilient quantum computation: error models and thresholds.
\newblock {\em Proc. R. Soc. Lond. A}, 454:365--384, 1998,
  \href{http://www.arxiv.org/abs/quant-ph/9702058}{{\tt
  arXiv:quant-ph/9702058}}.

\bibitem[Kni04]{Knill04schemes}
Emanuel Knill.
\newblock Fault-tolerant postselected quantum computation: schemes, 2004,
  \href{http://www.arxiv.org/abs/quant-ph/0402171}{{\tt
  arXiv:quant-ph/0402171}}.

\bibitem[Kni05]{Knill05}
Emanuel Knill.
\newblock Quantum computing with realistically noisy devices.
\newblock {\em Nature}, 434:39--44, 2005.

\bibitem[PV08]{PlenioVirmani08magic}
Martin~B. Plenio and Shashank~S. Virmani.
\newblock Upper bounds on fault tolerance thresholds of noisy {C}lifford-based
  quantum computers, 2008, \href{http://www.arxiv.org/abs/0810.4340}{{\tt
  arXiv:0810.4340 [quant-ph]}}.

\bibitem[Raz04]{Razborov}
Alexander~A. Razborov.
\newblock An upper bound on the threshold quantum decoherence rate.
\newblock {\em Quant. Inf. Comput.}, 4(3):222--228, 2004,
  \href{http://www.arxiv.org/abs/quant-ph/0310136}{{\tt
  arXiv:quant-ph/0310136}}.

\bibitem[Rei05]{Reichardt04magic}
Ben~W. Reichardt.
\newblock Improved magic states distillation for quantum universality.
\newblock {\em Quant. Inf. Proc.}, 4:251--264, 2005,
  \href{http://www.arxiv.org/abs/quant-ph/0411036}{{\tt
  arXiv:quant-ph/0411036}}.

\bibitem[Rei06a]{Reichardt06thesis}
Ben~W. Reichardt.
\newblock {\em Error-detection-based quantum fault tolerance against discrete
  {P}auli noise}.
\newblock PhD thesis, University of California, Berkeley, 2006,
  \href{http://www.arxiv.org/abs/quant-ph/0612004}{{\tt
  arXiv:quant-ph/0612004}}.

\bibitem[Rei06b]{Reichardt05distancethree}
Ben~W. Reichardt.
\newblock Fault-tolerance threshold for a distance-three quantum code.
\newblock In {\em Proc. Int. Conf. on Automata, Languages and Programming
  (ICALP)}, LNCS 4051, pages 50--61, 2006,
  \href{http://www.arxiv.org/abs/quant-ph/0509203}{{\tt
  arXiv:quant-ph/0509203}}.

\bibitem[Rei09]{Reichardt07algorithmica}
Ben~W. Reichardt.
\newblock Error-detection-based quantum fault-tolerance threshold.
\newblock {\em Algorithmica}, 55(3):517--556, 2009.

\bibitem[Shi03]{Shi02}
Yaoyun Shi.
\newblock Both {T}offoli and {C}ontrolled-{NOT} need little help to do
  universal quantum computation.
\newblock {\em Quant. Inf. Comput.}, 3(1):84--92, 2003,
  \href{http://www.arxiv.org/abs/quant-ph/0205115}{{\tt
  arXiv:quant-ph/0205115}}.

\bibitem[Ste03]{Steane03}
Andrew~M. Steane.
\newblock Overhead and noise threshold of fault-tolerant quantum error
  correction.
\newblock {\em Phys. Rev. A}, 68:042322, 2003,
  \href{http://www.arxiv.org/abs/quant-ph/0207119}{{\tt
  arXiv:quant-ph/0207119}}.

\bibitem[VHP05]{VirmaniHuelgaPlenio}
Shashank~S. Virmani, Susana~F. Huelga, and Martin~B. Plenio.
\newblock Classical simulability, entanglement breaking, and quantum
  computation thresholds.
\newblock {\em Phys. Rev. A}, 71:042328, 2005,
  \href{http://www.arxiv.org/abs/quant-ph/0408076}{{\tt
  arXiv:quant-ph/0408076}}.

\end{thebibliography}

\end{document}